\documentclass[11pt]{article}

\usepackage[utf8]{inputenc}
\usepackage[T1]{fontenc}
\usepackage{lmodern}
\usepackage{todonotes}
\usepackage{graphicx}
\usepackage{amsmath,amssymb}
\usepackage{booktabs}
\usepackage{siunitx}
\usepackage[font=small,labelfont=bf]{caption}
\usepackage{authblk}
\usepackage{xcolor}
\usepackage{hyperref}
\hypersetup{colorlinks=true,linkcolor=blue,citecolor=blue,urlcolor=blue}
\usepackage[margin=1in]{geometry}
\usepackage{placeins}
\usepackage[most]{tcolorbox}   

\graphicspath{{figures/}}

\title{\textbf{HClimRep-Ocean: A Global Ocean Emulator on an Unstructured Mesh}}

\author[1]{Kacper Nowak}
\author[1]{Aleksei Koldunov}
\author[1]{Nikolay Koldunov}
\author[2]{Savvas Melidonis}
\author[2]{Ankit Patnala}
\author[2]{Simon Grasse}
\author[6]{Julius Polz}
\author[4]{Christian Lessig}
\author[2,3]{Martin Schultz}
\author[1,5]{Thomas Jung}
\affil[1]{Alfred Wegener Institute, Helmholtz Centre for Polar and Marine Research, Bremerhaven, Germany}
\affil[2]{Forschungszentrum Jülich GmbH, Jülich Supercomputing Center, Jülich, Germany}
\affil[3]{University of Cologne, Department of Mathematics and Computer Science, Cologne, Germany}
\affil[4]{European Center for Medium-Range Weather Forecasts, Bonn, Germany}
\affil[5]{Department of Physics and Electrical Engineering, University of Bremen, Bremen, Germany}
\affil[6]{Karlsruhe Institute of Technology, Karlsruhe, Germany}

\date{\today}

\begin{document}
\maketitle

\begin{abstract}
Machine-learning (ML) emulators for atmospheric processes have advanced rapidly in recent years, transforming weather forecasting. Although early ML ocean forecasting models now exist, they remain less developed than their atmospheric counterparts. Unlike the atmosphere, much of the ocean's kinetic energy resides in mesoscale eddies whose characteristic spatial scales are approximately an order of magnitude smaller than those of comparable atmospheric features. Moreover, complex coastlines, narrow straits, and ice-covered seas make boundary representation a central challenge that atmospheric models do not face. Consequently, numerical ocean simulations commonly use locally refined or even completely unstructured meshes. However, their data-driven counterparts have so far been built around latitude-longitude grids. We present HClimRep-Ocean, an ocean emulator that operates directly on the native unstructured mesh of FESOM2. The emulator is trained on a 209-year AWI-CM3 control integration and is run without atmospheric forcing, receiving the atmospheric state only at initialisation time, which isolates the predictability carried by the ocean state itself.  Skill is strongly field-dependent: for currents, HClimRep-Ocean outperforms every reference at 30 day forecast, whereas for temperature and salinity a damped-anomaly persistence forecast remains the more accurate estimator. This behaviour is physically interpretable: current variability is largely geostrophic and internally generated, whereas sea-surface temperature and salinity fluctuations are driven by atmospheric forcing through weather state. Evaluated independently on the OceanBench benchmark, a reanalysis-trained variant of HClimRep-Ocean achieves the lowest RMSE against GLORYS reanalysis among all assessed systems, confirming the competitiveness of the native-mesh approach.
\end{abstract}


\section{Introduction}
\label{sec:intro}

Over the past few years, data-driven machine-learning (ML) models have catalysed a
shift in Earth-system modeling. In numerical weather prediction (NWP),
autoregressive forecast models such as GraphCast \cite{lam2023graphcast}, the
generative ensemble model GenCast \cite{price2025gencast} and ECMWF's Artificial
Intelligence Forecasting System (AIFS) \cite{lang2024aifs} now match or exceed
operational physics-based models across many variables, at a small fraction of the
computational cost. Many of these systems are built on graph neural networks (GNNs)
that pass messages over a learned mesh, an architecture that has become a de-facto
standard for modeling global geophysical fields and now underpins shared operational
frameworks such as ECMWF's Anemoi \cite{lang2024aifs}. Once trained, such emulators
produce a global forecast in near real-time, shifting the main computational burden to the
training phase and removing the need to integrate discretised equations of motion at
forecast time \cite{bi2023pangu, pathak2022fourcastnet}.

Skillful ocean forecasts underpin marine services ranging from navigation and marine safety to fisheries, offshore operations and hazard response; more fundamentally, the ocean is the slow component and the primary memory of the coupled climate system, and its thermal inertia and large-scale circulation make it a dominant source of predictability on sub-seasonal-to-seasonal (S2S) timescales, well beyond the roughly two-week deterministic horizon of the chaotic atmosphere \cite{s2s}. Efficient data-driven ocean emulators are therefore valuable both for accelerating ocean forecasting and for laying the groundwork toward fully coupled, data-driven Earth-system models. Yet ocean emulation also poses distinctive challenges. Unlike atmospheric processes, much of the ocean's kinetic energy resides in mesoscale eddies. Their characteristic scale—as set by the Rossby radius of deformation—is roughly an order of magnitude smaller than the atmospheric counterpart and contracts markedly toward the poles \cite{wenhai, rosby}. Furthermore, the ocean is bounded by complex coastlines, threaded by narrow straits and marginal seas, and overlain by seasonal ice. These constraints have no analogue in the atmosphere, meaning that energy-containing scales are turbulent, sharply localized, and intrinsically more difficult to resolve and model.

Several global ocean ML systems have nonetheless emerged in quick succession. One family of models targets short- to medium-range forecasting trained on ocean reanalysis: XiHe \cite{xihe}, the first data-driven $1/12^\circ$ eddy-resolving global system, uses a hierarchical transformer trained on the GLORYS12 reanalysis \cite{GLORYS12}. WenHai \cite{wenhai} reaches comparable skill while incorporating air--sea bulk formulae driven by ERA5 atmospheric fields to better preserve mesoscale eddy variability. Mercator Ocean's GLONET \cite{glonet} delivers operational $\sim$10-day global forecasts, and AI-GOMS \cite{aigoms}, trained on the HYCOM reanalysis, extends deterministic forecasts to 30 days while providing a reusable backbone for downstream tasks. More recently, FuXi-Ocean \cite{fuxiocean} and TianHai \cite{tianhai} have pushed global prediction to sub-daily (six-hourly) frequency at eddy-resolving resolution, the latter again coupling the ocean to a data-driven atmosphere. The second model family targets climate timescales by training on the output of physics-based ocean models: Samudra \cite{samudra} emulates a state-of-the-art ocean model across the full water column and remains stable for centuries. Alongside these three-dimensional systems, a parallel line of work  learns surface fields alone, chiefly sea-surface temperature, to provide lower boundary
conditions for atmospheric models in sub-seasonal-to-seasonal prediction without representing the interior ocean \cite{aimip}. The growing number of systems has prompted community-level intercomparison efforts such as OceanBench \cite{oceanbench}, which evaluates diverse models against shared references.

Two limitations recur across this body of work and motivate the present study. The first limitation is dynamical. Deterministic autoregressive emulators tend to
dampen high-frequency and small-scale variability. Trained under mean-squared-error
objectives, they relax toward the conditional mean and progressively smooth the
unpredictable mesoscale field as lead time grows. This tendency is well documented.
WenHai's central design choice of ingesting air--sea fluxes is motivated precisely by
the need to counteract it \cite{wenhai} --- and it points to the role of continued
atmospheric forcing in sustaining fast ocean variability. The second limitation is
geometric. The systems above operate on regular latitude--longitude grids, which
distribute degrees of freedom uniformly and therefore over-resolve the quiescent open
ocean while under-resolving the dynamically active, geometrically constrained regions
that matter most: coastlines, narrow straits, shelf seas, marginal ice zones and
energetic western boundary currents. 

Modern ocean general-circulation models avoid this compromise by using unstructured, 
variable-resolution meshes that concentrate resolution where the
physics demands it \cite{foxkemper2019}. Training an ML emulator on such output
currently requires interpolating the native fields onto a regular grid, which
discards that adaptive resolution and introduces systematic projection artifacts and
conservation errors, particularly near complex coastlines \cite{gridartifacts}.
Recent work has begun to apply graph-based ML to ocean domains with complex geometry,
but not yet on the native mesh of a global ocean model. SeaCast \cite{seacast} applies
hierarchical GNNs to the Mediterranean Sea, but builds its graph from an underlying
regular $1/24^\circ$ grid and is forced by the atmosphere at every step. The global
probabilistic emulator Njord \cite{njord} likewise uses a GNN, and pointedly operates
only over ocean points rather than a full latitude--longitude grid, but constructs its
graph by spherical $k$-means clustering of a regular $0.25^\circ$ field rather than by
adopting a physical model's own discretisation. To our knowledge, no system runs
directly on the parent ocean model's native unstructured computational mesh.
This is a missed opportunity: meshes such as the FESOM2 D3 configuration concentrate 
resolution down to $\sim$3\,km in western boundary currents, and the NG5 mesh 
reaches $\sim$5\,km globally, representing the state of the art in unstructured ocean 
modeling and the natural target for mesh-native emulation.

In this study we address both the geometric bottleneck and the variability-damping
problem with a transformer-based ocean emulator. Throughout this paper, ``emulator'' means an ML
model trained to reproduce the output of a specific physics-based model aiming to
imitate that model's own trajectory rather than to forecast the real ocean. 
Our model, named HClimRep-Ocean\footnote{The name alludes to the HClimRep project, part of the Helmholtz Foundation Model Initiative (HFMI) of Helmholtz Association; we gratefully acknowledge their support.}, 
operates directly on the native unstructured FESOM2 mesh \cite{danilov2017fesom2,scholz2019fesom2}, 
working with the model's cell--vertex topology directly and thereby eliminating ocean-side
regridding. We further run it forcing-free after initialisation, supplying the
atmospheric state only at the initial time. This is a deliberate baseline rather than
an omission: by withholding future forcing we isolate the predictability carried by
the ocean state itself and obtain a controlled setting in which to quantify how much
short-timescale variability depends on ongoing air--sea coupling. 
Because the emulator receives no atmospheric input after
initialisation, we expect that atmospherically driven
surface fields (sea-surface temperature, salinity) lose predictive power relative to a
forced system. This is not a deficiency of the mesh or the architecture but the
expected behaviour under the stochastic-climate framework of
Hasselmann~(1976)~\cite{hasselmann1976}: we report this loss quantitatively and
interpret it as evidence of the field-dependent structure of intrinsic ocean
predictability. Re-introducing atmospheric forcings is planned as an extension of this
work.

Beyond the 30-day verification set, we also integrate the emulator autoregressively for
180 days from a single initial condition. With one year of validation data remaining, only
one such rollout is available, so this serves as a qualitative stability check rather than a 
skill assessment; it nevertheless shows that the model remains bounded and preserves 
the mean stratification far beyond the range at which it retains deterministic skill.
Additionally, a reanalysis-trained variant of the same forcing-free architecture has been evaluated 
on the OceanBench intercomparison \cite{oceanbench} against five operational and data-driven forecasting systems,
achieving the lowest root-mean-square error relative to the GLORYS reanalysis across
the majority of variables, depths and lead times (Section~\ref{sec:oceanbench}). This
independent assessment confirms both the competitiveness of the native-mesh approach
and the physical consistency inherent in the forcing-free design.


\section{Data and model}
\label{sec:data}

\subsection{Training}
\label{sec:awicm}

In this work, the training data consists of simulated data from the third version of the coupled Alfred Wegener Institute Climate Model (AWI-CM3 v3.1.3) \cite{streffing2022awicm3}. Its ocean component is the second version of the FESOM2 on the CORE2 mesh \cite{scholz2019fesom2}. The CORE2 mesh is a standard global unstructured grid for FESOM2, featuring a nominal $1^\circ$ horizontal resolution with refinement down to roughly 20-25 km in the tropics and high latitudes (this accounts for approximately $126,858$ surface nodes, and $244,659$ triangle elements \cite{scholz2019fesom2}).
Throughout this paper, \emph{nodes} denote the FESOM2
grid vertices at which scalar fields (temperature, salinity, sea-surface height)
are defined, and \emph{elements} denote the triangle centroids that carry the
horizontal velocity components. The same graph-based architecture extends without
modification to strongly refined meshes such as the FESOM2 D3 configuration
($\sim$3\,km in western boundary currents) and the NG5 mesh ($\sim$5\,km
globally), which represent the state of the art in unstructured ocean modeling.

\begin{table}[htbp]
    \centering
    \caption{Typical horizontal resolution of the CORE2 mesh.}
    \label{tab:core2-resolution}
    \begin{tabular}{l c}
        \toprule
        Region                               & Nominal resolution \\
        \midrule
    High latitudes (poleward of $60^\circ$) & $20$ km \\
        Tropics ($30^\circ$ S–$30^\circ$ N)      & $25$ km \\
        Mid‑latitudes ($30^\circ$–$60^\circ$)    & $35$ km \\
        Remaining ocean                       & $50-120$ km \\
        \bottomrule
    \end{tabular}
\end{table}

Specifically, the data source is a
$\sim$210-year IFS--FESOM (AWICM3) control run with fixed (non-transient)
greenhouse-gas forcing. However, the
coupling is one-directional for the purpose of the emulator: the ocean fields
used for training are shaped by the atmosphere, but the emulator itself receives
no atmospheric input beyond the initial condition (Section~\ref{sec:arch}), and
the ocean exerts no feedback on the atmospheric component during emulator
inference.
Prognostic fields are potential temperature and salinity on \num{20} vertical levels, 
sea-surface height, sea-surface salinity and sea-surface temperature at nodes,
and horizontal velocity $(u,v)$ on elements. The 20 retained levels lie at
2, 7, 15, 25, 35, 45, 55, 65, 75, 85, 95, 107, 125, 147, 175, 210, 255, 310, 375
and 450\,m; 450\,m is the deepest level retained by the emulator. 

The atmospheric state supplied at initialization consists of 9 fields taken from the OpenIFS
component of the same AWI-CM3 integration, so that atmosphere and ocean come from a single 
coupled trajectory. They span the channels through which the atmosphere acts on the upper ocean:
the two 10\,m wind components, which set the surface stress; 2\,m temperature and dewpoint temperature,
which together determine the sensible and latent heat fluxes; convective and large-scale precipitation,
separated because they differ in spatial coherence; net top-of-atmosphere shortwave radiation under 
all-sky and clear-sky conditions, whose difference encodes the cloud shading of the surface; 
and mean sea-level pressure, which provides the inverse-barometer loading on sea-surface height. 

\paragraph{Training configuration.}
Training uses 209 model years (years 2000--2208) with a temporal resolution of
24\,h, i.e.\ daily snapshots. The final year (2209) is held out for validation.
In order to maximize the training set and rely on the 329 daily
initializations from the held-out year to provide robust skill statistics
(Section~\ref{sec:methods}), training ran for approximately \num{33000} optimizer
steps on samples drawn randomly from the training period. Further details on the
training configuration, including the loss function and learning-rate schedule,
are given in Appendix~\ref{app:training}. 

\paragraph{OceanBench variant}
A separate instance of the same forcing-free architecture was pre-trained on
64~years of the EERIE IFS--FESOM high-resolution coupled integration
\cite{eeriedata} and subsequently fine-tuned on the GLORYS12 ocean reanalysis
\cite{GLORYS12} with ERA5 atmospheric fields \cite{era5} for independent
evaluation on the OceanBench intercomparison
(Section~\ref{sec:oceanbench}). 
Both the EERIE fields and the GLORYS12 fields were conservatively remapped
to a regular $1/4^\circ$ grid using CDO \cite{cdo} before training.
However, ERA5 was ingested on native N320 reduced Gaussian grid. The
variant therefore operates partially on a structured grid, but is otherwise 
identical in architecture and inference protocol.
Because it differs only in its training data and input grid, we describe its
results alongside the main model but do not detail its training procedure
further here.

\paragraph{Reference-run drift and internal variability.}
Because the source is a control run with fixed (non-transient) greenhouse-gas forcing, it carries no prescribed
external-forcing trend; the only secular changes are the coupled model's own
control drift, which is weak over the archived 210 model years. The global-mean
sea-surface temperature drifts at $+2.6\,\mathrm{mK\,yr^{-1}}$, sea-surface
salinity at $-0.55\,\mathrm{m\,psu\,yr^{-1}}$, and surface kinetic energy shows no
significant trend ($\approx3\%$ interannual spread); the fastest-drifting
diagnostic is the amplitude (spatial standard deviation) of the dynamic topography,
at $+0.34\,\mathrm{mm\,yr^{-1}}$ on a $\sim$0.7\,m base, reflecting the slow spin-up
of the circulation. Projected onto the longest rollout considered here (180 days, \S\ref{sec:methods}) these amount to
$\le$1.5\,mK, $\le$0.3\,m\,psu and $\le$0.2\,mm respectively --- negligible relative
to the errors reported in \S\ref{sec:results} --- so drift of the verification
target does not contaminate the skill estimates. The internal variability is the
model's own: the validation year (2209) is climatologically ordinary, its El
Ni\~no--Southern Oscillation, Atlantic Multidecadal and North Atlantic Oscillation
indices all lying within $\pm1\sigma$ of the control-run distribution
($-0.9\sigma$, $+0.0\sigma$, $-0.5\sigma$), so the reported skill is not conditioned
on an anomalous ocean state.

\subsection{Model Architecture}

\begin{figure}
    \centering
    \includegraphics[width=1\linewidth]{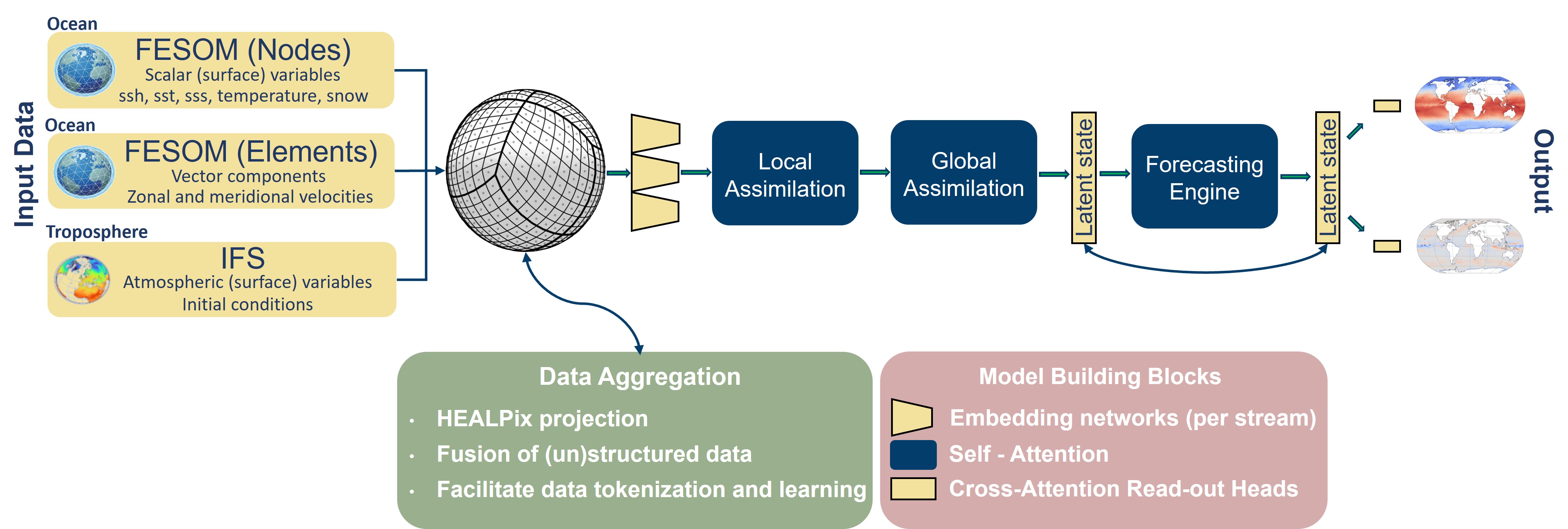}
    \caption{The HClimRep-Ocean model makes use of the \href{https://github.com/ecmwf/WeatherGenerator}{WeatherGenerator} model prototype which comprises of a series of attention-based transformer blocks. There are three main groups (engines) of such blocks: the local assimilation engine, the global assimilation engine, and the forecasting engine. Each of those engines attends to a different level of the model’s latent representation: the local assimilation engine attends within each cell to combine embedded input tokens from different streams into a fixed number of latent vectors per cell; then the global assimilation engine, by using dense attention over local neighborhoods on the sphere which are defined through the HEALPix subdivision, processes these vectors across cells combining them in a latent Earth system state; the forecasting engine attends to this global state to advance the temporal window by temporal window $\Delta t$.}
    \label{fig:prototype}
\end{figure}

\label{sec:arch}
\paragraph{HClimRep-Ocean.} 
HClimRep-Ocean is built on the deterministic variant of the WeatherGenerator
prototype \cite{ecmwf_weathergenerator}, an attention-based architecture
\cite{vaswani2017attention} originally developed for medium-range weather
forecasting. Two adaptations distinguish HClimRep-Ocean from the parent model
(see Figure~\ref{fig:prototype}): first, it ingests ocean fields on the native
FESOM2 unstructured mesh together with atmospheric fields supplied only at
initialisation (Section~\ref{sec:awicm}); second, it targets forecast ranges well
beyond the medium-range weather horizon, up to 30\,days for quantitative
verification and 180\,days for stability assessment. The architecture consists of
four stages --- encoding, assimilation, latent-space forecasting and decoding ---
whose configuration details are given in Appendix~\ref{app:model_architecture}.

\paragraph{Encoding.} The embedding and encoding stage maps the input data to a learned
representation. Inputs are organized into streams, where each stream is associated with a dedicated transformer-based embedding network. We use three streams: two streams for the surface‑node channels and element channels of the ocean component respectively, and one stream for the IFS atmospheric near‑surface channels, see Figure \ref{fig:prototype}. Before being processed by the embedding networks, the data points are spatially organized on a HEALPix grid. Each data point is assigned to its corresponding HEALPix cell based on its geographical location, and data points that share the
same cell are grouped together. When a cell contains more inputs than a specified threshold, they are redistributed across multiple cells. The HEALPix resolution is chosen to match the information density of the data.

\paragraph{Assimilation.} The assimilation engine consists of three components: the local
assimilation engine, read‑out heads, and a global assimilation engine. The local assimilation engine computes attention between embedded tokens from the different input streams for each cell using a dense transformer. Inspired by Perceiver‑IO mechanism \cite{jaegle2021perceiver}, read-out heads then use 
cross-attention to project these tokens onto a fixed-size representation per cell: a small set of learned, cell-shared read-out tokens act as queries attending to them as keys and values, mapping each cell into a shared global latent space. Applying this projection for every cell yields a fixed number of latent vectors, which are processed by the global assimilation engine, a transformer with dense self-attention that combines local information into a globally consistent latent state. The sparse attention operates on local neighborhoods on the sphere, conveniently defined through the HEALPix scheme.

\paragraph{Latent Space Forecasting.} The forecasting engine is implemented by a transformer with dense self‑attention, analogous to the global assimilation engine. Given a latent state from the global assimilation engine, associated with a temporal window, it generates latent states for subsequent temporal windows, progressively extending the forecast.

\paragraph{Decoding.} Decoding from the latent space back to physical space is handled by dedicated projection heads, which mirror the read-out step in reverse: per-stream target coordinates act as queries attending to the per-cell latent states, producing field values at the target coordinates for each stream, as illustrated in Figure~\ref{fig:prototype}.


\section{Verification methodology}
\label{sec:methods}

\paragraph{Forecast sets.}
Two forecast sets are used: 329 daily initialisations of the held-out year integrated for
30 days, which carry all quantitative skill results, and a single 180-day integration from
one initial condition, used for the long-rollout stability assessment in
\S\ref{sec:horizon}.

\paragraph{Verification target.}
By construction the emulator is a surrogate for the AWI-CM3/FESOM2 coupled model, so its
only well-posed verification target is the parent model's own trajectory. Comparing
against observations or reanalysis would conflate two distinct error sources: the
emulator's fidelity to AWI-CM3, and AWI-CM3's own departure from the real ocean. The
latter is a property of the physics-based model, not of the emulator. We therefore verify
strictly against the parent model. Benchmarks such as OceanBench \cite{oceanbench} and the
model-to-observations protocols used by reanalysis-trained systems (XiHe, WenHai, GLONET)
address a different question, which is operational skill against the real ocean and are out
of scope for an emulator evaluated against its source model. We note, however, that a
reanalysis-trained variant of the same architecture has been independently evaluated on
OceanBench (\S\ref{sec:oceanbench}), providing a complementary assessment against
observational references; incorporating observations into the emulator's own training is a
separate line of work (\S\ref{sec:limitations}).

\paragraph{Baselines.}
We compare against (i) persistence, the initial ocean state (the lead-1
target field) held fixed and (ii) a day-of-year climatology built from a
30-year window of AWI-CM3 (model years 2180--2209). The verification year lies inside
that window and therefore contributes one thirtieth of the climatological mean, which
marginally favours the climatology baseline; we retain the full window for consistency
with the archived climatology. Two caveats follow. First,
persistence anchored at the initial state is trivially exact at the first lead and
is therefore a strong competitor at very short range, so the model
overtakes it only after the first week or two rather than at lead zero. Second,
climatology is a demanding baseline when the verification year drifts relative to
the climatology window; we therefore treat persistence as the primary skill
reference for the seasonal set.
 
As a third, more stringent baseline we use \emph{damped-anomaly persistence}:
the MSE-optimal linear combination of persistence and climatology, whose weight
is the lag autocorrelation \cite{murphy1992}. In the stochastic-climate framework of
Hasselmann~\cite{hasselmann1976} and Frankignoul and Hasselmann~\cite{frankignoul1977},
upper-ocean temperature anomalies obey a first-order autoregressive process driven by
atmospheric weather noise. In case of setting such noise to zero, as is the case 
for our forcing-free emulator yields pure exponential decay toward climatology.
This baseline is therefore the theoretically expected behaviour of the forcing-free 
emulator for atmospherically driven fields, and beating it is a necessary condition 
for the emulator to add value beyond a trivial relaxation of the initial anomaly. The forecast is defined as
\begin{equation}
  \hat{x}(t+L) = \mathrm{clim}(\mathrm{doy}) + \alpha(L)\,[x(t_0) - \mathrm{clim}],
\end{equation}
with an exponential damping $\alpha(L)=e^{-L/\tau}$. The $e$-folding time $\tau$ is
selected per field from $\{3,5,7,10,15,20,25,30\}$ days as the value that minimizes
the baseline's own area-weighted error over the verification set, giving $\tau=30$\,d
for temperature and salinity and $\tau=15$\,d for sea-surface height and currents.
Because $\tau$ is tuned in-sample, and tuned to the baseline's advantage, this is a
deliberately hard-to-beat reference.

\paragraph{Metrics.}
All spatial statistics are weighted by the true FESOM2 node and element areas,
the correct weighting on a mesh whose cell areas vary by more than two orders of
magnitude. We use root-mean-square error (RMSE), anomaly correlation (ACC), normalised
standard deviation, and skill scores $S = 1 - \mathrm{RMSE}_\mathrm{model}/
\mathrm{RMSE}_\mathrm{baseline}$. Anomalies are taken about the day-of-year climatology
throughout, except for the variance diagnostics, which use anomalies about the local
time-mean of the target. Temporal variability is assessed with area-weighted power
spectra, computed per node on linearly detrended 30-day forecast series and averaged
over 24 initializations, two per calendar month.

\paragraph{Regions and depths.}
We stratify by three zonal bands (NH extratropics, $>20^\circ$N; tropics,
$\pm20^\circ$; SH extratropics, $>20^\circ$S) and six dynamically active regions
--- the Gulf Stream, Kuroshio, Agulhas, Brazil--Malvinas confluence, tropical band
and Southern Ocean --- and report depth profiles to 450\,m; the regional breakdown is given in Appendix~\ref{sec:regions}.

\section{Results}
\label{sec:results}

\subsection{Global skill versus lead time}
\label{sec:skill}
\begin{table}
\centering
\caption{Forecast accuracy of the FESOM2 emulator at the 15- and 30-day horizons.
Root-mean-square errors and standard deviations are area-weighted global means over wet mesh points, verified
against the FESOM2 reference simulation. Skill is measured against a persistence
forecast that carries the initial state forward, and is formed for each initialisation
before averaging rather than as a ratio of the averaged errors. Uncertainties are the
standard deviation across the 329 initialisations.}
\label{tab:scores}
\begin{tabular}{lcccc}
\toprule
 & \multicolumn{2}{c}{RMSE} & \multicolumn{2}{c}{Skill vs pers.} \\
\cmidrule(lr){2-3}\cmidrule(lr){4-5}
Field & +15 d & +30 d & +15 d & +30 d \\
\midrule
SST & $0.61{\pm}0.07$\,$^\circ$C & $0.69{\pm}0.08$\,$^\circ$C & $0.30{\pm}0.12$ & $0.50{\pm}0.11$ \\
SSS & $0.25{\pm}0.01$\,psu & $0.31{\pm}0.02$\,psu & $0.08{\pm}0.11$ & $0.20{\pm}0.11$ \\
SSH & $0.045{\pm}0.005$\,m & $0.050{\pm}0.005$\,m & $0.20{\pm}0.07$ & $0.18{\pm}0.07$ \\
\addlinespace
$T$ @ 25\,m & $0.55{\pm}0.03$\,$^\circ$C & $0.65{\pm}0.04$\,$^\circ$C & $0.23{\pm}0.11$ & $0.41{\pm}0.12$ \\
$T$ @ 95\,m & $0.37{\pm}0.01$\,$^\circ$C & $0.46{\pm}0.02$\,$^\circ$C & $0.31{\pm}0.04$ & $0.37{\pm}0.04$ \\
$T$ @ 210\,m & $0.279{\pm}0.005$\,$^\circ$C & $0.349{\pm}0.012$\,$^\circ$C & $-0.02{\pm}0.08$ & $0.03{\pm}0.06$ \\
$T$ @ 450\,m & $0.204{\pm}0.004$\,$^\circ$C & $0.242{\pm}0.006$\,$^\circ$C & $-0.65{\pm}0.06$ & $-0.39{\pm}0.03$ \\
\addlinespace
$S$ @ 25\,m & $0.182{\pm}0.008$\,psu & $0.219{\pm}0.009$\,psu & $-0.09{\pm}0.04$ & $0.04{\pm}0.05$ \\
$S$ @ 95\,m & $0.093{\pm}0.003$\,psu & $0.103{\pm}0.004$\,psu & $-0.58{\pm}0.14$ & $-0.29{\pm}0.11$ \\
$S$ @ 210\,m & $0.073{\pm}0.001$\,psu & $0.079{\pm}0.001$\,psu & $-1.72{\pm}0.18$ & $-1.13{\pm}0.13$ \\
$S$ @ 450\,m & $0.069{\pm}0.001$\,psu & $0.074{\pm}0.004$\,psu & $-4.67{\pm}0.21$ & $-3.15{\pm}0.19$ \\
\addlinespace
$|U|$ @ 2.5\,m & $0.105{\pm}0.006$\,m\,s$^{-1}$ & $0.115{\pm}0.006$\,m\,s$^{-1}$ & $0.40{\pm}0.04$ & $0.39{\pm}0.03$ \\
$|U|$ @ 25\,m & $0.071{\pm}0.004$\,m\,s$^{-1}$ & $0.083{\pm}0.004$\,m\,s$^{-1}$ & $0.48{\pm}0.03$ & $0.45{\pm}0.03$ \\
$|U|$ @ 95\,m & $0.039{\pm}0.002$\,m\,s$^{-1}$ & $0.052{\pm}0.004$\,m\,s$^{-1}$ & $0.63{\pm}0.02$ & $0.53{\pm}0.03$ \\
$|U|$ @ 210\,m & $0.0287{\pm}0.0006$\,m\,s$^{-1}$ & $0.0361{\pm}0.0008$\,m\,s$^{-1}$ & $0.55{\pm}0.02$ & $0.49{\pm}0.01$ \\
$|U|$ @ 450\,m & $0.0239{\pm}0.0004$\,m\,s$^{-1}$ & $0.0299{\pm}0.0005$\,m\,s$^{-1}$ & $0.55{\pm}0.02$ & $0.50{\pm}0.01$ \\
\bottomrule
\end{tabular}
\end{table}

\begin{figure}
    \centering
    \includegraphics[width=0.9\linewidth]{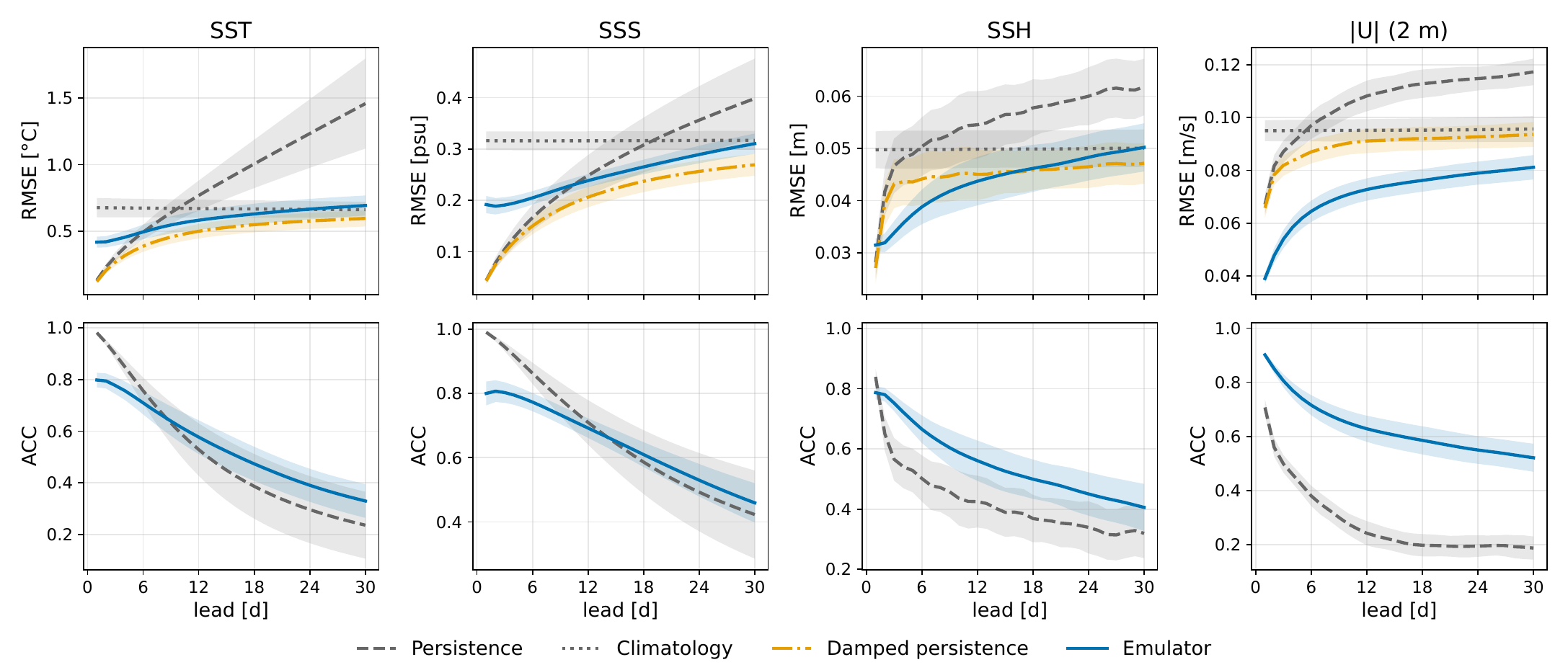}
    \caption{Surface forecast skill of the emulator against three references,
    versus lead time (mean over 329 daily initializations of the validation year;
    shading shows $\pm1\sigma$ across those initializations).
    Columns: sea-surface temperature (SST), salinity (SSS), height (SSH) and current speed $|U|$
    at 2\,m; top row root-mean-square error (RMSE, weighted by the FESOM2 node and element areas), bottom row
    anomaly correlation coefficient (ACC) relative to the 30-year day-of-year climatology.
    References are persistence, damped persistence and climatology.
    The emulator beats raw persistence and climatology at all but the shortest leads, but only
    the currents clearly beat the damped-persistence
    baseline; for SST and SSS the emulator does not, and for SSH only marginally near two weeks.}
    \label{fig:surf_skill}
\end{figure}

We verify the emulator against the FESOM2 reference simulation over the held-out year 2209,
using every daily initialization of that year ($N=329$) and integrating each forecast
forward for 30 days. 
Table~\ref{tab:scores} summarises the scores at the 15- and 30-day horizons and
Fig.~\ref{fig:surf_skill} shows their full dependence on lead time.

Surface errors grow rapidly during the first days and then flatten markedly. Sea-surface
temperature errors reach $0.42\,^\circ$C after one day, $0.48\,^\circ$C after five and
$0.61\,^\circ$C at 15 days, so that 88\,\% of the eventual 30-day error
($0.70\,^\circ$C) has already accumulated by the middle of the rollout. This saturating
behaviour is characteristic of a forecast that is progressively relaxing towards the
model's own attractor rather than diverging: the anomaly correlation decays steadily from
$0.80$ at one day to $0.52$ at 15 days and $0.33$ at 30 days, while the error rises
towards the climatological level, which it reaches at day 24. The emulator overtakes
persistence after 7 days for SST, after 2 days for sea-surface height and after 11 days
for sea-surface salinity, the ordering reflecting the intrinsic persistence timescale of
each field. Measured against climatology, SST retains useful information until day 24 and SSH until
day 30, whereas SSS remains more accurate than climatology throughout the 30-day window.

Damped persistence is a far more demanding reference, and it separates the thermodynamic
from the dynamic fields. The fitted damping itself already distinguishes them: for surface
temperature $\alpha$ falls from $0.98$ at one day to $0.60$ at 15 days and $0.43$ at
30 days, and for salinity from $0.99$ to $0.71$ and $0.56$, whereas for surface velocity it
collapses from $0.69$ to $0.13$ within a fortnight. Anomalies of temperature and salinity
are long-lived, so a damped copy of the initial anomaly remains an excellent forecast of
them; velocity anomalies decorrelate within days, so dampened persistence there is barely
better than climatology.

Against this reference the emulator loses for the thermodynamic surface fields and wins
for the dynamic ones. Surface temperature error is $0.613\,^\circ$C at 15 days against
$0.532\,^\circ$C for damped persistence, and $0.697$ against $0.598\,^\circ$C at 30 days;
surface salinity is $0.252$ against $0.222$\,psu and $0.311$ against $0.263$\,psu at the
same horizons. Dampened persistence is more accurate than the emulator for these two fields
at every lead of the rollout, by $14$--$18$\,\%. Sea-surface height is marginal: the
emulator is ahead between days 2 and 15 and behind thereafter, but by less than one per
cent at two weeks ($0.0454$ against $0.0457$\,m at 15 days, $0.050$ against
$0.047$\,m at 30 days), so the two are effectively indistinguishable over the first
fortnight. Horizontal velocity is the
one field in which the emulator is clearly ahead: at 2.5\,m it improves on dampened
persistence by $18$\,\% at 15 days and $11$\,\% at 30 days, and at 95\,m by $48$\,\% and
$31$\,\%, at every lead of the rollout.

This field-dependent pattern is a direct consequence of the forcing-free experimental
design and is predicted by the stochastic-climate framework invoked in
\S\ref{sec:methods}. With atmospheric forcing removed after initialization, the emulator's
surface tracer fields evolve without the noise that sustains their anomalies in the coupled
model; the resulting exponential decay toward climatology is the theoretically expected
behavior, and damped persistence is its statistical expression. Where the reference
simulation's anomalies persist over the forecast range, the emulator's step-by-step
integration accumulates error faster than its learned dynamics recover information, and a
simple exponential relaxation of the initial anomaly is the better estimator. For
currents, whose variability is dominated by geostrophic adjustment rather than
atmospheric weather noise, anomalies decorrelate within days and no such statistical
shortcut exists; here the learned dynamics carry the forecast. The comparison also shows
that the gains over plain persistence reported above are in part a statement about the
weakness of that baseline for slowly evolving fields rather than about the emulator alone.
We regard damped persistence as the reference of record for the surface tracer fields, and
the emulator's ability to beat it for currents as evidence that it has learned intrinsic
ocean dynamics beyond a trivial relaxation. Moreover, whereas damped persistence produces
an independent scalar forecast at each grid point with no spatial consistency, the emulator
generates dynamically coherent ocean states in which temperature, salinity, and velocity
fields evolve together, a property that is invisible in point-wise scores but essential
for applications that require spatially consistent fields.

The subsurface behavior differs qualitatively from the surface and is best read through
the anomaly correlation rather than the raw error. Absolute errors fall with depth simply
because variability does, but the correlation with the reference simulation is markedly
\emph{higher} below the surface: temperature at 95\,m retains an anomaly correlation of
$0.84$ at 15 days and $0.74$ at 30 days, against $0.52$ and $0.33$ at the surface, and
even at 450\,m it stays above the surface value ($0.68$ and $0.57$). Horizontal velocity
is the best-predicted field at every level, with component-mean anomaly correlations of
$0.87$ at 15 days and $0.76$ at 30 days at 95\,m. Velocity also improves on persistence from the
first forecast day onward at all depths, with skill scores of $0.4$--$0.6$, consistent
with a field whose short decorrelation time makes persistence a weak baseline. The
surface layer is therefore not where the emulator is most skilful in a relative sense; it
is where the flow it has to reproduce is most strongly and most rapidly forced.

Table~\ref{tab:scores} also exposes two weaknesses. Deep temperature is not improved
upon persistence: at 450\,m the skill score is $-0.65$ at 15 days, meaning that simply
holding the initial state fixed would be more accurate than integrating the emulator,
even though the forecast still beats climatology at every lead. In a water mass whose
anomalies persist for months, the model's incremental updates add more error than signal.
Second, and more seriously, subsurface salinity is degraded relative to both references.
Below the mixed layer the emulator is already worse than climatology from the first day
at 450\,m and from day 18 at 95\,m, and its skill against persistence reaches $-4.7$ at
450\,m. Deep salinity anomalies are weak, long-lived and only loosely coupled to the
surface fields that dominate the training loss, so the network has little incentive to
preserve them; the practical consequence is that the present configuration should not be
used to advect deep salinity structure, and that depth-dependent loss weighting is the
natural remedy to test.

\begin{figure}[t]
\centering
\includegraphics[width=\textwidth]{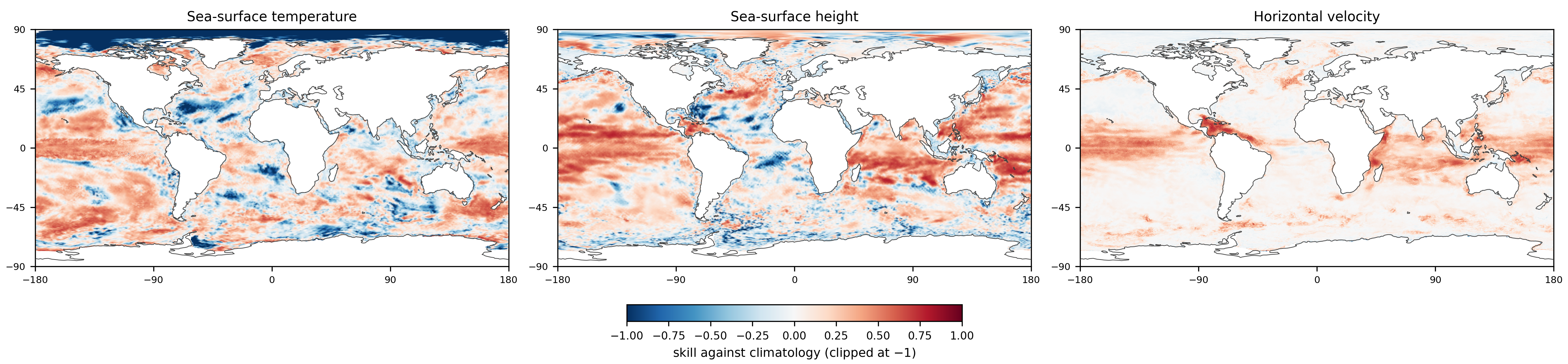}
\caption{Where the forecast beats the seasonal cycle. Each panel maps the skill against
the day-of-year climatology, $1-\mathrm{RMSE}/\mathrm{RMSE}_{\mathrm{clim}}$, evaluated at
every mesh point over the $N=329$ held-out initialisations at the 15-day horizon; red
denotes a forecast more accurate than climatology, blue less accurate and white equally
accurate, and the scale is clipped at $-1$. Velocity skill is computed from the vector
error, $\mathrm{RMSE}_{|U|}^{2}=\mathrm{RMSE}_{u}^{2}+\mathrm{RMSE}_{v}^{2}$, applied to
both the forecast and the climatology. The layout matches Fig.~\ref{fig:variability} so
that the two may be compared directly: that figure shows where the forecast carries the
right amount of variability, this one where it is actually more accurate than the seasonal
cycle. Skill is broadly positive in the tropics and subtropics for all three fields and
negative over the Arctic and parts of the subpolar gyres for surface temperature.}
\label{fig:skill_maps}
\end{figure}

The global figures also conceal a pronounced geographical structure, which
Fig.~\ref{fig:skill_maps} resolves by mapping the same skill against climatology at each
mesh point. Two features stand out. First, skill is not distributed as the error is: the
tropics, where absolute errors are modest, are where the emulator gains most on
climatology (mean skill $+0.10$ for surface temperature, $+0.24$ for sea-surface height
and $+0.21$ for surface velocity), while the Arctic is the one region where surface
temperature is decisively worse than the seasonal cycle (mean skill $-0.82$), the seasonal
ice edge being both highly variable and only weakly constrained by the initial state.
Second, the fields differ in how widely rather than how strongly they are skilful: surface
temperature beats climatology over $63\,\%$ of the ocean and velocity over $81\,\%$, yet
their global mean skills are similar, because velocity's gains are broad and shallow while
temperature's are large in some regions and strongly negative in others. Averaged scores of
the kind reported above therefore describe the tropics and subtropics well and the polar
oceans poorly, and the regional breakdown should be consulted before transferring these
numbers to a high-latitude application.

Finally, the global means conceal a systematic hemispheric asymmetry. Northern-hemisphere
SST errors exceed southern-hemisphere ones by roughly 50\,\% at 15 days ($0.74$ against
$0.49\,^\circ$C) and the same contrast holds for salinity ($0.31$ against $0.19$\,psu),
reflecting the stronger mesoscale activity and deeper wintertime mixing of the northern
basins. The seasonal and regional structure of this contrast is examined in
Sect.~\ref{sec:season}.

\FloatBarrier
\subsection{Vertical structure}
\label{sec:depth}

\begin{figure}
    \centering
    \includegraphics[width=\linewidth]{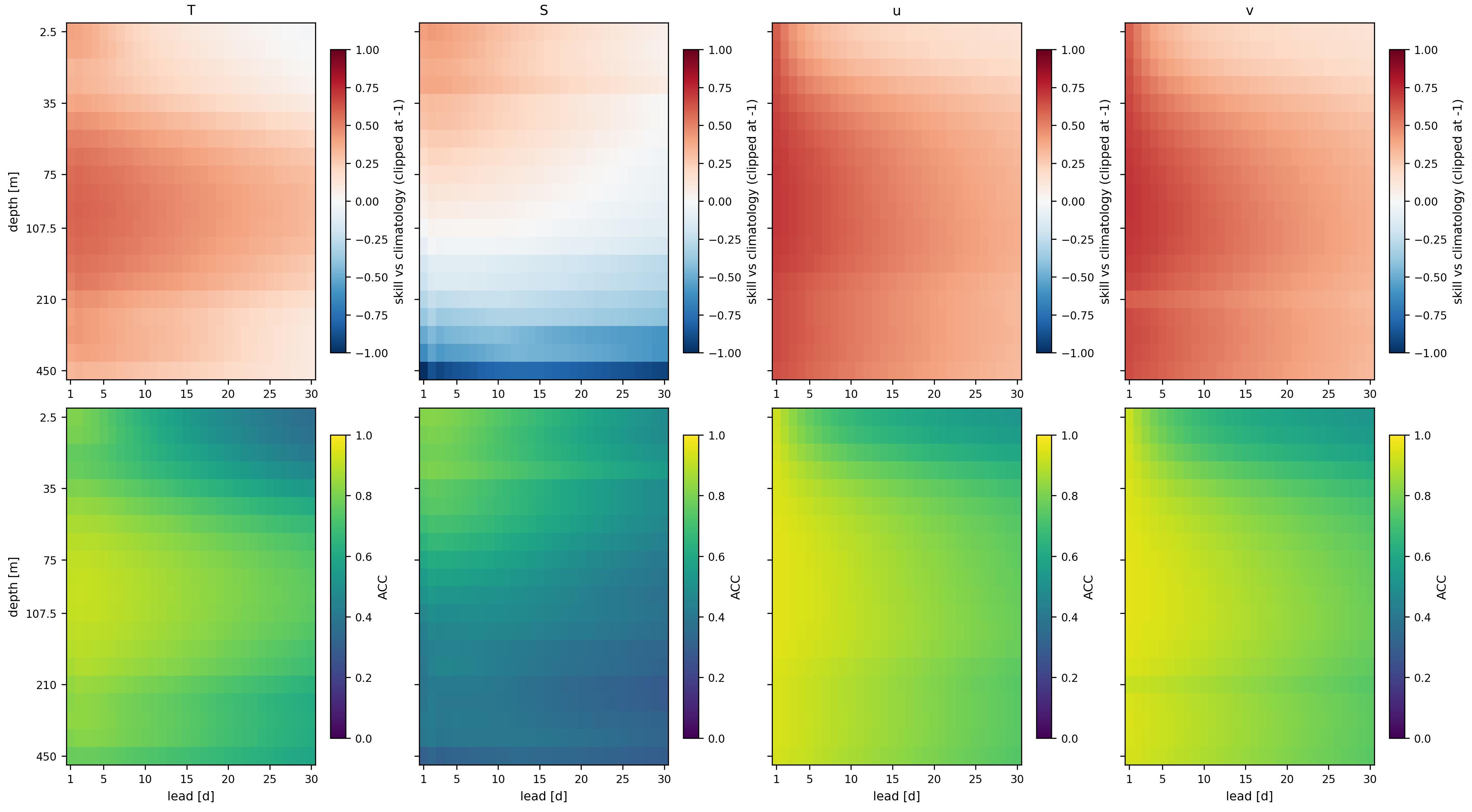}
    \caption{Forecast skill as a joint function of lead time and depth, for temperature,
        salinity and the two horizontal velocity components (columns). The upper row shows the
        skill against the day-of-year climatology,
        $1-\mathrm{RMSE}/\mathrm{RMSE}_{\mathrm{clim}}$, so that positive values indicate a
        forecast more accurate than climatology, zero indicates equal accuracy and negative
        values indicate that climatology would have been the better estimate; the scale is
        clipped at $-1$. The lower row shows the anomaly correlation on a fixed $0$--$1$ scale.
    }

    \label{fig:lead_depth}
\end{figure}

\begin{figure}
    \centering
    \includegraphics[width=\linewidth]{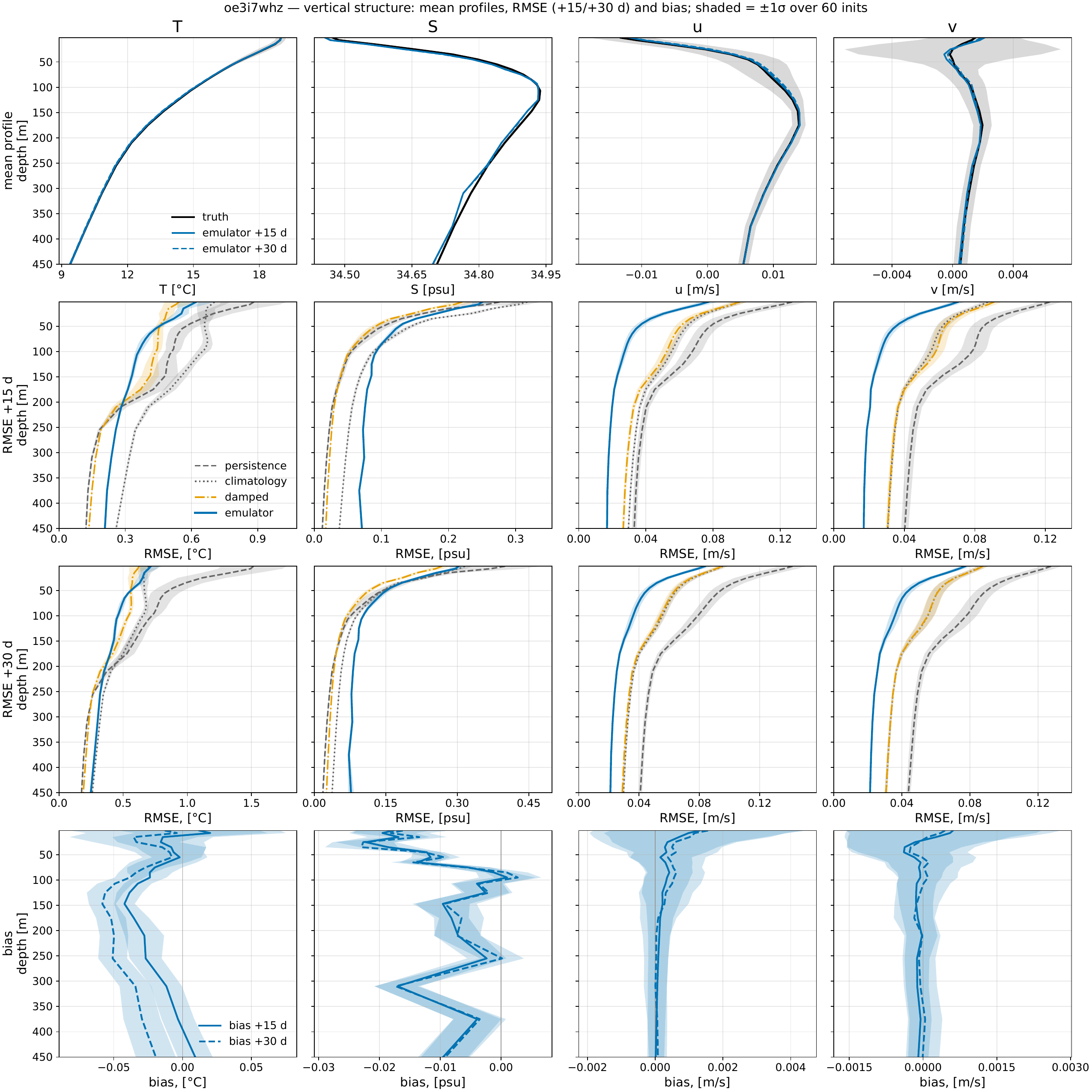}
    \caption{Vertical structure of the forecast, for temperature, salinity and the two
  horizontal velocity components (columns). Top row: area-weighted mean profiles of the
  reference simulation and of the emulator at +15 and +30\,d. Second and third rows:
  root-mean-square error against depth at +15 and +30\,d, for the emulator and the three
  baselines. Bottom row: mean error (bias, emulator $-$ reference) at the same two leads.
  Shading shows $\pm1\sigma$ across 60 initialisations spanning the held-out year;
  statistics use the FESOM2 node and element areas and exclude cells below the local
  bathymetry. The mean stratification is preserved to +30\,d, and the emulator improves on
  every baseline for the velocity components at all depths, but below roughly 100\,m at
  +15\,d --- and below 35\,m at +30\,d --- its salinity error exceeds that of climatology.}
    \label{fig:profiles}
\end{figure}

Figure~\ref{fig:profiles} summarizes how the forecast behaves through the water
column. For temperature, salinity, and the two horizontal velocity components it shows the
area-weighted mean profile of the reference simulation together with the emulator's
profiles at 15 and 30 days (top row), the root-mean-square error against depth (middle
row) and the mean error, or bias, against depth (bottom row). The complementary view of
the same information as a function of both depth and lead time is given in
Fig.~\ref{fig:lead_depth}.

The mean stratification is preserved. Over a 30-day rollout the global mean temperature
profile drifts by at most $0.05\,^\circ$C, and the mean salinity profile by at most
$0.02$\,psu, at every level between the surface and 450\,m; on the scale of the top row
of Fig.~\ref{fig:profiles} the forecast and reference curves are visually
indistinguishable. These drifts are one to two orders of magnitude smaller than the
corresponding random errors, so the emulator does not accumulate a systematic warming,
cooling or freshening of the upper ocean over the forecast range considered here. This is
a non-trivial property for an autoregressive model integrated for thirty consecutive
steps without any relaxation to the reference state.

Random errors decrease monotonically with depth and closely track the vertical
distribution of variability. Temperature errors fall from $0.61\,^\circ$C at the surface
to $0.37\,^\circ$C at 95\,m and $0.20\,^\circ$C at 450\,m at the 15-day horizon, and
velocity errors fall by a factor of four over the same interval, from
$0.105$ to $0.024$\,m\,s$^{-1}$. The decline is smooth rather than peaked: we find no
local error maximum at the base of the mixed layer, which would be the signature of a
misplaced thermocline. Because variability decreases with depth at least as fast as the
error does, the \emph{relative} accuracy improves downwards, and the anomaly correlations
reported in Sect.~\ref{sec:skill} are correspondingly higher below the surface than at
it.

The bias profiles reveal the one coherent systematic signal in the upper ocean. Below
about 15\,m the emulator is systematically too cold, with the bias deepening from
$-0.01\,^\circ$C just below the surface to a maximum of $-0.043\,^\circ$C at 15 days and
$-0.061\,^\circ$C at 30 days near 150\,m, before relaxing again towards zero at 450\,m.
The signal is small in absolute terms: at the depth where it peaks it accounts for 14\,\%
of the total error there, and given the local mean temperature gradient it corresponds to
displacing the isotherms downwards by roughly two metres. Salinity shows the mirror-image
pattern in the surface layer, where the forecast is too fresh by about $0.02$\,psu, with
the bias passing through zero near 90\,m. Errors in the interior are therefore dominated
by their random component rather than by drift, but the growth of the thermocline cold
bias with lead time indicates a slow, coherent adjustment of the upper thermocline that
would become the leading error source in longer integrations.

Read together with Sect.~\ref{sec:skill}, the profiles explain why the emulator loses to
persistence at depth while still beating climatology: the deep ocean errors are small in
absolute terms and largely unbiased, but they are not small compared with the very slow
evolution of the deep fields themselves, so a forecast that simply holds the initial
state fixed is difficult to improve upon. The exception remains subsurface salinity,
whose error at 450\,m is $1.8$ times its own climatological spread and which the profiles
show to carry a persistent fresh bias at every depth below 300\,m.

\FloatBarrier
\subsection{Dependence on initialization season}
\label{sec:season}

The scores discussed so far average over a full year of initializations and therefore
conceal a pronounced seasonal cycle. We group the initializations by the
meteorological season of their start date and evaluate each hemisphere separately, since
the local season is reversed south of the equator. Table~\ref{tab:season}
collects the results, ordered by \emph{local} season so that each row compares the two
hemispheres in the same physical regime; the held-out year contains no December
initializations, so the DJF sample covers January and February only.

Surface temperature error is governed by the local season rather than by the calendar.
It grows by a factor of $1.9$ between local winter and local summer initializations in
the northern hemisphere, and by a factor of $1.8$ in the southern hemisphere: reading
down the two SST columns of Table~\ref{tab:season}, the same ordering
(winter $<$ spring $\approx$ autumn $<$ summer) appears in both, although the calendar
seasons that produce it are six months apart. Anomaly correlations follow suit, falling
to $0.43$ at 15 days for northern summer starts against $0.59$ for northern winter ones.
That the two hemispheres show the same seasonal amplitude in antiphase is the clearest
available indication that the signal is physical rather than an artifact of the calendar
or of the training sample. The interpretation is the familiar one: a shallow summer mixed
layer has little thermal inertia and responds quickly to atmospheric forcing, so surface
temperature anomalies are both larger and shorter-lived, whereas the deep winter mixed
layer damps and retains them. Surface salinity behaves in the same way, with error maxima
for local summer starts in both hemispheres ($0.37$ against $0.27$\,psu in the north,
$0.24$ against $0.14$\,psu in the south).

Skill relative to persistence peaks in a different season, and again does so
symmetrically: the largest gain occurs for local autumn initializations in both
hemispheres, about twice the value obtained in the other seasons and rising further by 30
days (to $0.64$ in the north and $0.61$ in the south). Autumn is when the mixed layer
deepens and re-entrains the anomalies accumulated over summer, so the surface state
evolves rapidly away from its initial condition and a persistence forecast degrades
quickly. This is precisely the regime in which an emulator that carries the dynamics adds
the most value, and conversely the reason why its advantage is smallest in winter, when
persistence is already an excellent forecast. Note that the seasons of largest error and
of largest skill therefore do not coincide: absolute accuracy is highest in winter, but
the forecast is most \emph{useful} in autumn.

Sea-surface height reverses the phase of the cycle. Its error is largest for local winter
initializations in both hemispheres and smallest in local summer, the opposite ordering to
temperature, with anomaly correlations correspondingly lowest in winter ($0.41$ for
northern DJF starts). Winter sea level is dominated by the storm-driven and convective
response of the high-latitude basins, which is both more energetic and less predictable
than the summer state. The seasonal cycles of thermal and dynamic surface error are thus
not merely different in amplitude but opposite in sign, which also means that no single
season is uniformly favourable for the model.

Two practical consequences follow. First, a verification campaign restricted to a single
season can misstate surface temperature error by nearly a factor of two, and would rank
the model against persistence quite differently depending on the season chosen; annual,
hemisphere-resolved verification is necessary for a meaningful headline number. Second,
the seasonal modulation of skill is not a defect to be tuned away but a property inherited
from the reference simulation, and reproducing it in both hemispheres is itself evidence
that the emulator has learned the seasonal reorganisation of the upper ocean rather than a
single climatological regime.

\begin{table}[t]
\centering
\caption{Seasonal dependence of forecast error at the 15-day horizon, ordered by
\emph{local} season so that each row compares the two hemispheres in the same
physical season; the calendar seasons that this corresponds to are given in
parentheses (northern / southern) and $N$ is the number of initialisations
contributing to each entry. Scores are area-weighted means within each hemisphere,
with skill measured against persistence as in
Table~\ref{tab:scores}. Surface temperature error peaks for local
summer initializations and skill against persistence peaks for local autumn
initializations in both hemispheres, whereas sea-surface height error peaks in local
winter: the seasonal cycles are of the same shape in the two hemispheres and
opposite in sign between the thermal and the dynamic surface field.}
\label{tab:season}
\begin{tabular}{llcccccc}
\toprule
 & & \multicolumn{2}{c}{SST RMSE ($^\circ$C)} & \multicolumn{2}{c}{SST skill vs pers.} & \multicolumn{2}{c}{SSH RMSE (m)} \\
\cmidrule(lr){3-4}\cmidrule(lr){5-6}\cmidrule(lr){7-8}
Local season (N / S) & $N$ & NH & SH & NH & SH & NH & SH \\
\midrule
Winter (DJF / JJA) & 59 / 92 & $0.52$ & $0.37$ & $0.16$ & $0.23$ & $0.059$ & $0.048$ \\
Spring (MAM / SON) & 92 / 86 & $0.65$ & $0.47$ & $0.27$ & $0.23$ & $0.042$ & $0.042$ \\
Summer (JJA / DJF) & 92 / 59 & $0.99$ & $0.69$ & $0.25$ & $0.21$ & $0.037$ & $0.039$ \\
Autumn (SON / MAM) & 86 / 92 & $0.65$ & $0.47$ & $0.44$ & $0.42$ & $0.053$ & $0.044$ \\
\bottomrule
\end{tabular}
\end{table}

\FloatBarrier
\subsection{Variability and spectra}
\label{sec:variability}

\begin{figure}[t]
\centering
\includegraphics[width=\textwidth]{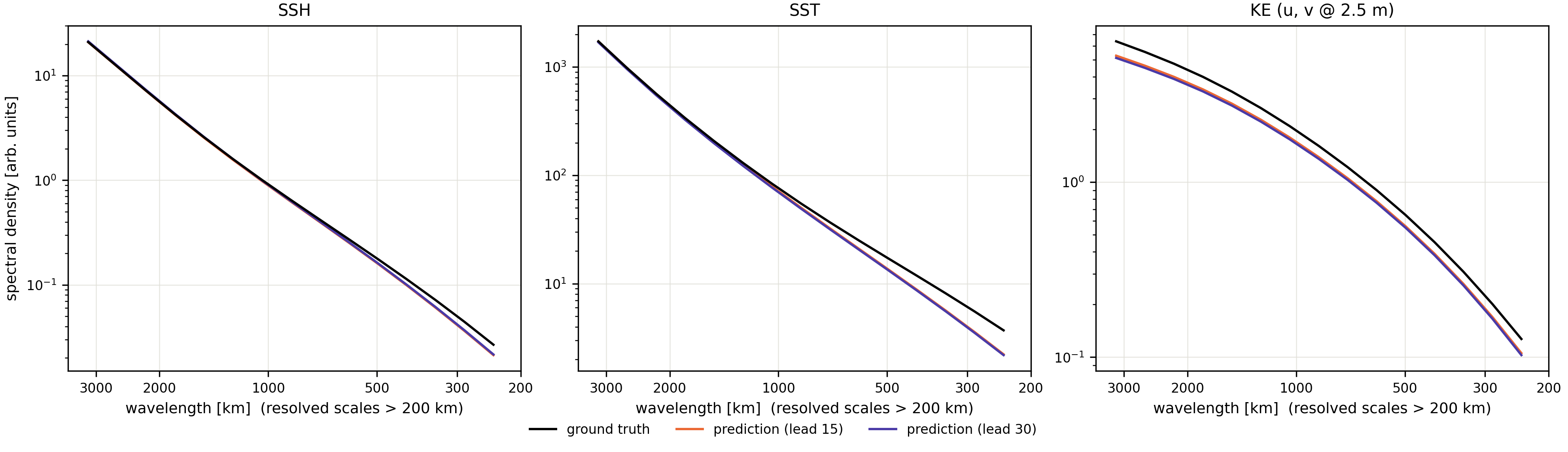}
\caption{Spatial variability of the forecast. Wavelength spectra computed directly on the
native triangular mesh by implicit filtering \cite{implicit_filter}, for sea-surface height, sea-surface
temperature and surface kinetic energy, comparing the reference simulation with the
emulator at the 15- and 30-day horizons; curves are averaged over the initialisations
sampled for this diagnostic. Wavelength decreases to the right and the axis stops at
200\,km, below which the CORE2 mesh no longer resolves the flow reliably. Forecast and
reference coincide at basin scales and separate towards smaller ones for the two tracer
fields, whereas kinetic energy is deficient by a roughly constant factor at every resolved
scale. The 15- and 30-day curves nearly overlap, indicating that the spectral deficit is
established early in the rollout.}
\label{fig:spectra}
\end{figure}

\begin{figure}[t]
  \centering
  \includegraphics[width=\linewidth]{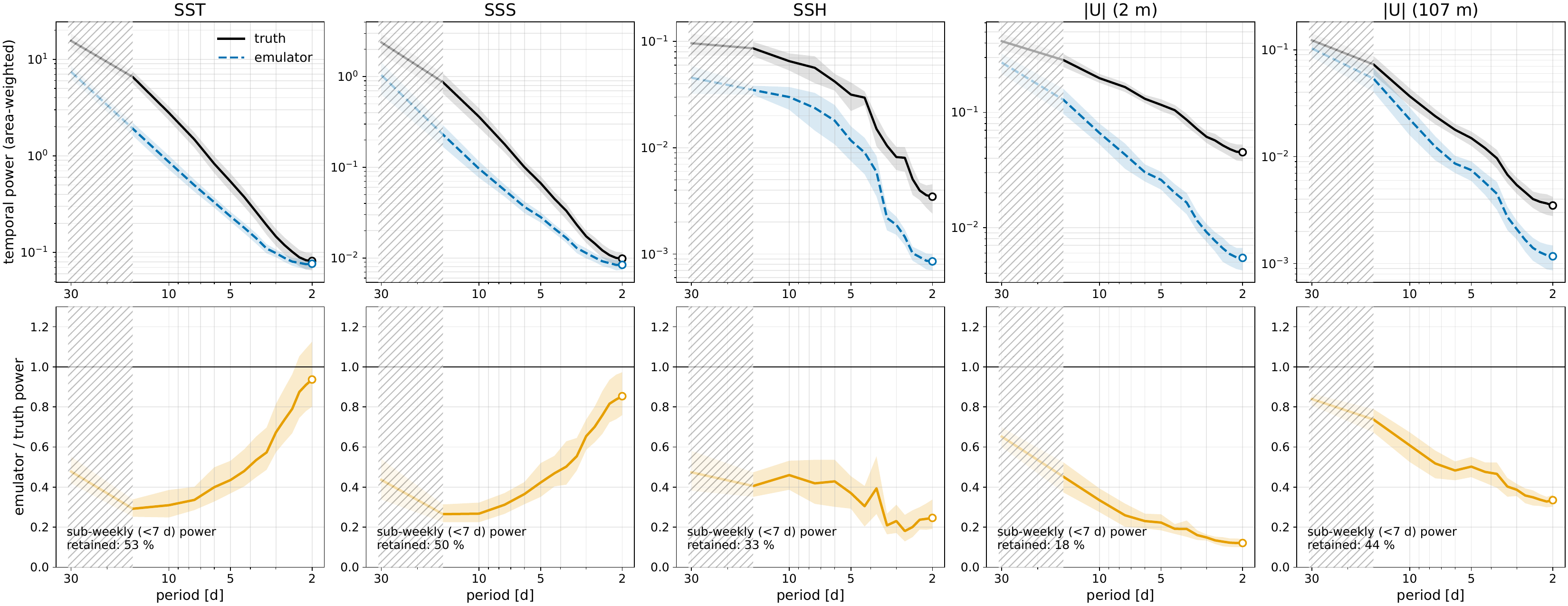}
  \caption{Temporal power spectra of the 30-day forecasts (top; truth solid, emulator dashed)
and the emulator/truth power ratio (bottom) for SST, SSS, SSH and current speed at 2\,m and
107\,m. Spectra are computed per node from linearly detrended 30-day lead series,
area-weighted over the mesh and averaged over 24 initialisations (two per calendar month), with shading showing $\pm1\sigma$ across those initialisations;
resolved periods are 2--30\,d. The hatched band (periods $\geq$15\,d, i.e.\ at most two
cycles in the record and affected by detrending) is qualitative only; open markers mark the
Nyquist period (2\,d). The percentage in each lower panel is the sub-weekly ($<$7\,d) power
retained by the emulator. For currents and SSH the emulator damps the short periods, most
strongly at the surface (18\,\% of sub-weekly power retained at 2\,m, 44\,\% at 107\,m,
33\,\% for SSH). For SST and SSS the loss within the 30-day window is broad-band and largest
at periods of 10--30\,d (ratio $\approx$0.3), reflecting the decay of anomaly amplitude rather
than a high-frequency cut-off; their sub-weekly power is retained at $\approx$50\,\%.}
  \label{fig:tspectra}
\end{figure}

Accuracy scores reward a forecast for being close to the reference in the mean square,
which a forecast can achieve by suppressing the variability it is unsure about. We
therefore examine how much variance the emulator actually carries, and where in scale and
frequency it is lost. Figure~\ref{fig:tspectra} shows the temporal power spectrum of each
field along the forecast trajectory, area-weighted and averaged over initialisations, with
the emulator-to-reference power ratio beneath; Fig.~\ref{fig:spectra} shows the
corresponding spatial spectra on the native mesh at the 15- and 30-day forecast horizons. Periods
longer than $\sim$ 14 days cannot be constrained by a 30-day rollout and are shaded in
Fig.~\ref{fig:tspectra}.

In time, the emulator is a low-pass filter. The reference and forecast spectra coincide at
the longest resolved periods and separate progressively towards higher frequencies, so that
of the power at periods shorter than a week the emulator retains $53\,\%$ for surface
temperature, $50\,\%$ for surface salinity, $44\,\%$ for velocity at 107\,m, $33\,\%$ for
sea-surface height and only $18\,\%$ for surface velocity. The ordering is physically
coherent: the fields whose sub-weekly variability is generated by rapid atmospheric forcing
and by the ageostrophic surface response are the ones most strongly damped, while the
slower, more balanced signals survive. The two tracers behave differently again at the very
shortest periods, where their power ratio recovers to $0.94$ and $0.86$ at the two-day
period, whereas surface velocity continues to fall to $0.13$; day-to-day tracer variability
is largely a direct thermodynamic response to the prescribed atmosphere, which the emulator
receives as input, while the corresponding velocity signal must be generated internally.

In time, the emulator is a low-pass filter. The reference and forecast spectra coincide at
the longest resolved periods and separate progressively towards higher frequencies, so that
of the power at periods shorter than a week the emulator retains $53\,\%$ for surface
temperature, $50\,\%$ for surface salinity, $44\,\%$ for velocity at 107\,m, $33\,\%$ for
sea-surface height and only $18\,\%$ for surface velocity. The ordering is physically
coherent: the fields whose sub-weekly variability is generated by rapid atmospheric forcing
and by the ageostrophic surface response are the ones most strongly damped, while the
slower, more balanced signals survive. This spectral hierarchy is the frequency-domain
expression of the stochastic-climate framework (\S\ref{sec:methods}): removing the
atmospheric noise source from a first-order ocean process suppresses variance at all
frequencies but preferentially at periods shorter than the ocean's intrinsic decorrelation
time, which is long for surface tracers and short for ageostrophic currents. The two
tracers behave differently again at the very shortest periods, where their power ratio
recovers to $0.94$ and $0.86$ at the two-day period, whereas surface velocity continues to
fall to $0.13$; day-to-day tracer variability is largely a direct thermodynamic response to
the prescribed atmosphere, which the emulator receives as input, while the corresponding
velocity signal must be generated internally.

\begin{figure}[t]
\centering
\includegraphics[width=\textwidth]{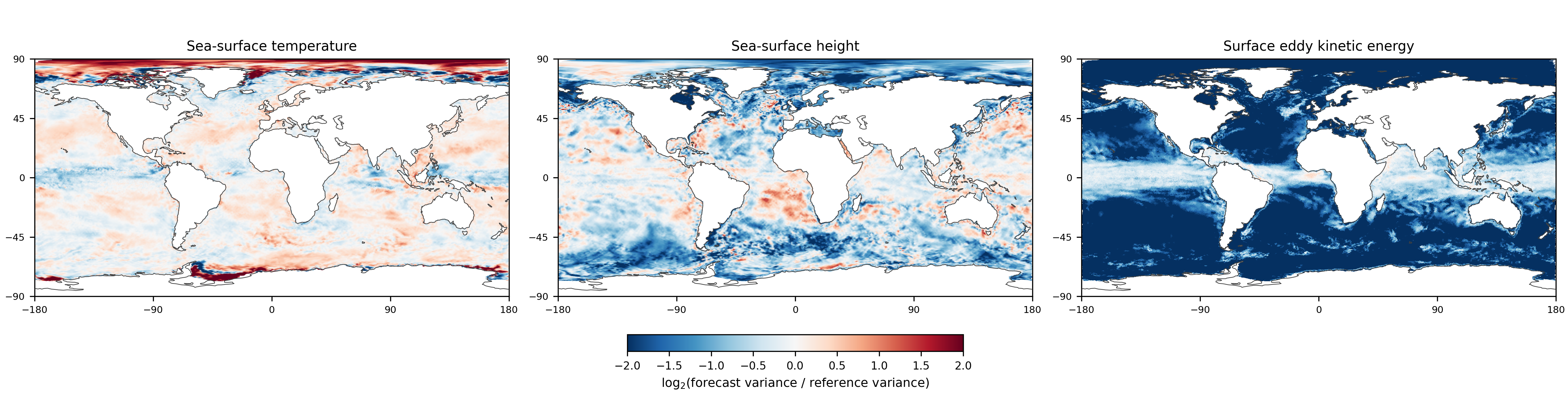}
\caption{Geographical distribution of the variance deficit. Each panel shows
$\log_2$ of the ratio between the forecast and reference variance across the all
initialisations at the 15-day horizon, so that zero denotes a forecast with the correct
amount of variability, negative values (blue) too little and positive values (red) too
much; eddy kinetic energy is formed from the two horizontal velocity components. Surface
temperature variance is close to correct and locally excessive in the western boundary
currents and the Southern Ocean, whereas eddy kinetic energy is deficient almost
everywhere outside the equatorial band, most severely in the Gulf Stream extension, the
Kuroshio, the Antarctic Circumpolar Current and the subtropical gyre interiors. The global
mean of this field is therefore dominated by the tropics and understates the deficit in the
regions where the mesoscale is most energetic.}
\label{fig:variability}
\end{figure}

The maps of forecast-to-reference variance ratio (Fig.~\ref{fig:variability}) show that
these two statements are not the same as a uniform loss of amplitude, and that the global
averages conceal a strong geographical structure. Taken across initializations at the
15-day horizon, the forecast
reproduces $104\,\%$ of the reference surface-temperature variance and $94\,\%$ of the
salinity variance at 15 days, but only $74\,\%$ for sea-surface height and $67\,\%$ for
surface eddy kinetic energy. The last figure is however dominated by the tropics, where
absolute velocity variance is large and the forecast retains $75\,\%$ of it; in the
eddy-rich regions the deficit is far more severe, with only $18\,\%$ of the surface eddy
kinetic energy retained in the Gulf Stream extension, $21\,\%$ in the Antarctic
Circumpolar Current, $26\,\%$ in the North Pacific interior and $34\,\%$ in the Kuroshio.
Surface temperature variance, by contrast, is if anything slightly too energetic in the
same western-boundary regions ($110\,\%$ and $113\,\%$ respectively). Total tracer variance
is therefore approximately conserved while being redistributed from fast, small scales
towards slow, large ones, whereas the dynamic fields lose variance outright and do so
precisely where the mesoscale is most active.

\FloatBarrier
\subsection{Predictability horizon and long-rollout behaviour}
\label{sec:horizon}

The results so far cover the 30-day range over which the emulator is intended to be used.
Integrated further, it may behave in qualitatively different ways, and which of them
applies cannot be read off a 30-day forecast. We therefore integrate the emulator for 180
days from a single initial condition. Only one such rollout is available, so this is a
qualitative assessment rather than a skill estimate; the possible behaviours are, however,
distinct enough to be told apart from a single realisation.

The three possibilities can be separated by a single identity. For any forecast, the error
relative to the day-of-year climatology is governed by just two numbers --- how well the
predicted anomalies still line up with the reference, and how large they are:
  \begin{equation}
    \frac{\mathrm{RMSE}}{\mathrm{RMSE}_{\mathrm{clim}}} = \sqrt{\,s^{2}-2\rho s+1\,},
    \label{eq:decomp}
  \end{equation}
where $\rho$ is the anomaly correlation with the reference and
$s=\sigma_{\mathrm{emulator}}/\sigma_{\mathrm{reference}}$ the ratio of anomaly amplitudes,
both taken about the day-of-year climatology. A forecast that collapses onto the
climatological mean loses its anomalies ($s\to0$) and its error tends to the climatological
value, unity. A forecast that keeps a realistic amplitude but loses all correlation ($s=1$,
$\rho=0$) settles at $\sqrt{2}$. An unstable integration is bounded by neither.
Equation~(\ref{eq:decomp}) also gives the lead at which a forecast ceases to beat
climatology, $\rho=s/2$, which reduces to $\rho=0.5$ when the amplitude is correct. Because
the emulator's amplitude departs from unity, this threshold moves: for surface currents,
whose anomalies are weaker than the reference, useful skill persists to a correlation
of $0.35$.

\begin{figure}[t]
  \centering
  \includegraphics[width=\textwidth]{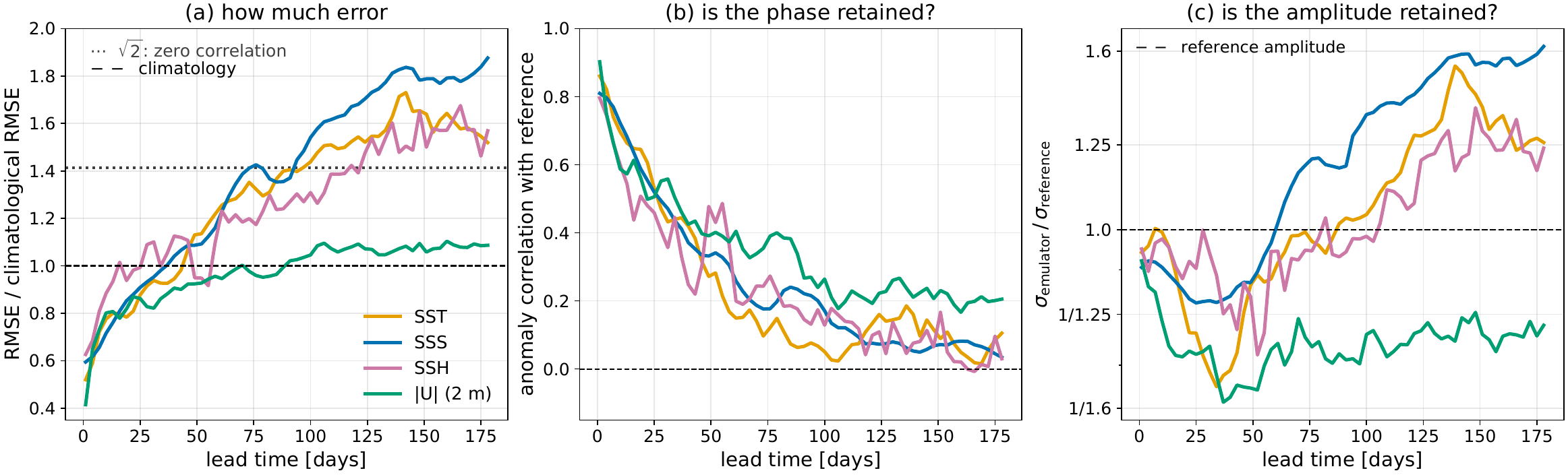}
  \caption{Behaviour of the emulator over a single 180-day rollout, decomposed through
  Eq.~(\ref{eq:decomp}). (a) Error relative to the day-of-year climatology; the dashed line
  marks parity with climatology and the dotted line the value $\sqrt{2}$ reached by a
  forecast that has lost all correlation with the reference while keeping its amplitude.
  (b) Anomaly correlation with the reference: the phase information. (c) Ratio of anomaly
  standard deviations: the amplitude, on a logarithmic axis so that equal factors above and
  below unity are equally far from it. Panels (b) and (c) together determine (a). All
  statistics are area-weighted with the FESOM2 node and element areas.}
  \label{fig:horizon}
\end{figure}

The emulator follows the second route (Fig.~\ref{fig:horizon}). Its error passes the
climatological level after 16 days for sea-surface height, 37 for salinity, 46 for surface
temperature and 70 for surface currents --- in each case within three days of the lead at
which $\rho$ falls below $s/2$, confirming that Eq.~(\ref{eq:decomp}) accounts for the
behaviour without a residual bias term. Thereafter the error saturates rather than
diverging: it reaches $1.5$--$1.9$ for the tracers and sea-surface height and only $1.09$
for surface currents, and its growth slows and, for surface temperature, stops altogether
after about 120 days. The anomaly correlation decays smoothly to $0.03$--$0.21$ over the
same period, so the phase information is essentially exhausted; the amplitude, however, is
not. It first drops to $0.65$--$0.78$ of the reference within the first month --- the
spectral damping of \S\ref{sec:variability} seen in the time domain --- and then, for the
tracers and sea-surface height, recovers and overshoots, reaching $1.24$--$1.62$ by day
180. Surface currents do not overshoot and remain at $0.78$. It is this excess amplitude,
not the loss of correlation, that carries the tracer curves above $\sqrt{2}$: at the same
correlations but a correct amplitude they would lie at $1.34$--$1.39$.

The mean state is preserved throughout. The global mean temperature and salinity profiles
at day 180 are indistinguishable from the reference on the scale of
Fig.~\ref{fig:profiles}, and the circulation remains energetic: the emulator retains
$86\,\%$ of the reference kinetic energy at 2\,m at day 180, against $66\,\%$ for the
climatology. That energy is however redistributed. Splitting it about the climatology, the
eddy component falls to $54\,\%$ of the reference at the surface but only to $83\,\%$ at
107\,m, so the flow is not slowed so much as smoothed --- the mean circulation survives
while the mesoscale is eroded, and preferentially where the wind-driven surface variability
lives. The long rollout therefore supports the conclusion drawn from the 30-day set: the
emulator does not relax onto a smooth mean state, which is the failure mode that would
matter most for climate-length integrations, but it reaches its stationary state with too
little mesoscale energy and, for the tracers, too much large-scale anomaly amplitude.

Two caveats attach to these numbers. They come from a single rollout, so the timing of
individual crossovers is uncertain. And the amplitude ratio is a quadratic statistic:
recomputed from robust (median-absolute-deviation) spreads, the surface-temperature
overshoot largely disappears ($1.04$ rather than $1.28$) while the current deficit deepens
($0.53$ rather than $0.78$), so the tracer overshoot is carried by a minority of points
whereas the current deficit is a property of the field as a whole.

\section{OceanBench Intercomparison}
\label{sec:oceanbench}
 
The results in Section~\ref{sec:results} evaluate the emulator strictly against
its parent coupled model. To assess whether the same architecture can also produce 
competitive forecasts we submitted a reanalysis-trained variant to the
OceanBench intercomparison \cite{oceanbench}. OceanBench is a community
benchmark maintained by Mercator Ocean International that evaluates global ocean
forecasting systems against the GLORYS12 reanalysis \cite{GLORYS12}, the GLO12
analysis, and in-situ observations over a shared verification period, using
root-mean-square deviation (RMSD) as the primary scalar metric. It differs from
the main model only in its training data: pre-training on 64~years of the EERIE
IFS--FESOM coupled integration \cite{eeriedata}, followed by fine-tuning on
GLORYS12 with ERA5 atmospheric fields \cite{era5,GLORYS12}
(Section~\ref{sec:data}). This retraining replaces the parent-model attractor
with the observed ocean state as the learning target, making verification against
reanalysis and observations a well-posed test of forecast skill rather than of
emulation fidelity.

Five other systems participated in the same evaluation round: GLO12, Mercator
Ocean's operational physics-based analysis--forecast system; GLONET
\cite{glonet}, Mercator Ocean's data-driven counterpart; WenHai \cite{wenhai},
which assimilates air--sea bulk fluxes at every forecast step; XiHe \cite{xihe},
a hierarchical-transformer system trained on GLORYS12; and LangYa, a
transformer-based system from IOCAS. All four data-driven challengers ingest
atmospheric forcing during the forecast, whereas HClimRep does not.

\paragraph{Comparison to the GLORYS reanalysis.}
Table~\ref{tab:oceanbench_surface} summarises the surface RMSD for all six
systems at lead days~1 and~10. For sea-surface temperature (SST), HClimRep
(0.57\,\textdegree C at day~1, 0.64\,\textdegree C at day~10) is competitive
with but slightly behind WenHai (0.55--0.61\,\textdegree C) and GLO12
(0.55--0.66\,\textdegree C). This is the expected cost of the forcing-free
design: without ongoing atmospheric input the emulator cannot track
atmospherically driven SST variability, the same mechanism discussed for the
parent-model emulator in Section~\ref{sec:results}. The gap is small, however,
and reverses at longer leads relative to GLO12, whose RMSD grows faster
(0.66\,\textdegree C at day~10 versus 0.64\,\textdegree C for HClimRep).
 
For all other surface variables, HClimRep achieves the lowest RMSD among all
challengers at every lead time evaluated. Sea-surface salinity is
0.41\,PSU at day~1 and 0.39\,PSU at day~10, roughly 20--40\,\% lower than the
next-best system (XiHe, 0.50--0.52\,PSU). Surface meridional and zonal current
RMSD is 0.11--0.13\,m\,s$^{-1}$, consistently below all forced systems
(0.13--0.17\,m\,s$^{-1}$). Sea-surface height RMSD is 0.06\,m across all lead
times, below GLO12 and WenHai (0.07--0.08\,m) and well below the remaining
data-driven systems (0.08--0.10\,m).
 
\begin{table}[htbp]
\centering
\caption{Surface RMSD against the GLORYS12 reanalysis for all OceanBench
challengers at lead days~1 and~10. Bold entries mark the lowest (best) value in
each column. Units: SST in \textdegree C, SSS in PSU, currents in
m\,s$^{-1}$, SSH in m. LangYa does not report day-10 scores.}
\label{tab:oceanbench_surface}
\small
\begin{tabular}{l cc cc cc cc cc}
\toprule
& \multicolumn{2}{c}{SST} & \multicolumn{2}{c}{SSS}
& \multicolumn{2}{c}{Merid.\ current} & \multicolumn{2}{c}{Zonal current}
& \multicolumn{2}{c}{SSH} \\
\cmidrule(lr){2-3}\cmidrule(lr){4-5}\cmidrule(lr){6-7}\cmidrule(lr){8-9}\cmidrule(lr){10-11}
Model & d1 & d10 & d1 & d10 & d1 & d10 & d1 & d10 & d1 & d10 \\
\midrule
HClimRep    & 0.57 & 0.64 & \textbf{0.41} & \textbf{0.39} & \textbf{0.11} & \textbf{0.13} & \textbf{0.11} & \textbf{0.13} & \textbf{0.06} & \textbf{0.06} \\
GLO12       & 0.55 & 0.66 & 0.63 & 0.64 & 0.13 & 0.16 & 0.13 & 0.17 & 0.07 & 0.08 \\
GLONET      & 0.67 & 0.91 & 0.56 & 0.56 & 0.13 & 0.15 & 0.13 & 0.15 & 0.08 & 0.10 \\
WenHai      & \textbf{0.55} & \textbf{0.61} & 0.59 & 0.58 & 0.13 & 0.14 & 0.13 & 0.15 & 0.07 & 0.08 \\
XiHe        & 0.63 & 0.79 & 0.52 & 0.50 & 0.13 & 0.13 & 0.14 & 0.14 & 0.08 & 0.09 \\
LangYa      & 0.68 & ---  & 0.62 & ---  & 0.13 & ---  & 0.14 & ---  & 0.07 & --- \\
\bottomrule
\end{tabular}
\end{table}

\paragraph{Subsurface fields.}
The advantage of HClimRep widens with depth. At 50\,m, temperature RMSD is
0.78--0.80\,\textdegree C across lead days~1--10, compared to 0.85--0.87\,\textdegree C
for the next-best system (WenHai) and 0.93--0.99\,\textdegree C for GLO12. At 100\,m
the gap grows further: HClimRep achieves 0.92--0.95\,\textdegree C while all other
systems exceed 1.0\,\textdegree C by day~10 (GLO12 reaches
1.17\,\textdegree C). At 200\,m, 300\,m and 500\,m, HClimRep retains the lowest
temperature RMSD among all challengers at every lead time, with values of
0.76--0.77, 0.63--0.64 and 0.48--0.49\,\textdegree C, respectively. Salinity
follows the same pattern: HClimRep is the lowest-error system at every depth from
the surface through 500\,m, with particularly large margins at the surface
(0.39\,PSU versus $\geq$0.50\,PSU for all others at day~10) and at 50\,m
(0.20\,PSU versus $\geq$0.25\,PSU). Subsurface current RMSD is likewise the
lowest or tied-lowest at 50\,m and 100\,m. Overall, HClimRep achieves the
lowest RMSD relative to the GLORYS reanalysis across the majority of
variable--depth--lead-time combinations, with sea-surface temperature the sole
variable at which a forced system (WenHai) scores lower.

\paragraph{Derived diagnostics.}
OceanBench additionally evaluates mixed-layer depth (MLD), surface geostrophic
currents and Lagrangian trajectory deviations. For Lagrangian trajectories,
HClimRep achieves the smallest deviation at every lead time: 9.7\,km at day~2,
growing to 67.5\,km at day~9, roughly 19\,\% below GLO12 (83.0\,km at day~9)
and 14--15\,\% below WenHai and GLONET. This is consistent with the low SSH and
current errors noted above, since Lagrangian displacement integrates the velocity
field over time. For geostrophic currents, HClimRep shows a distinctive pattern
of stable or slightly improving RMSD over the 10-day window (meridional:
0.18\,m\,s$^{-1}$ at day~1, decreasing to 0.16\,m\,s$^{-1}$ at day~10; zonal:
0.21 decreasing to 0.19\,m\,s$^{-1}$), while the forced data-driven systems
(GLONET, WenHai) degrade by a factor of two over the same period. This stability
is consistent with the emulator learning balanced dynamics whose geostrophic
component is tied to the slowly evolving SSH and density fields rather than to
atmospherically forced ageostrophic fluctuations that the forced systems must
track. Mixed-layer depth is the one diagnostic where GLO12 is clearly superior
(RMSD $\approx$32--36\,m versus $\approx$49--51\,m for HClimRep), reflecting
the importance of instantaneous surface buoyancy forcing in setting the MLD,
which the forcing-free emulator does not receive.
 
\paragraph{Comparison to in-situ observations.}
Against Argo-derived temperature and salinity profiles, all systems perform more
closely than against the reanalysis. At the surface (0--5\,m), SST RMSD ranges
from 0.76\,\textdegree C (WenHai) to 0.87\,\textdegree C (LangYa) at day~1;
HClimRep sits at 0.82\,\textdegree C, within $\sim$0.05\,\textdegree C of GLO12
(0.77\,\textdegree C). Near-surface salinity RMSD is 0.27\,PSU for HClimRep,
indistinguishable from most other systems. For near-surface currents (15\,m
depth, evaluated against drifter observations), HClimRep achieves the lowest
zonal-current RMSD at all lead times (0.20--0.22\,m\,s$^{-1}$, tied with GLONET)
and the lowest or tied-lowest meridional-current RMSD
(0.19--0.20\,m\,s$^{-1}$). Sea-level anomaly RMSD against altimetry is
0.06--0.07\,m, comparable to all other systems.

\paragraph{Interpretation.}
The OceanBench results confirm the pattern identified in the parent-model
evaluation: there is a tradeoff of modest penalty in atmospherically
driven surface temperature for a consistent advantage in dynamical fields (SSH,
currents, Lagrangian transport) and subsurface thermohaline structure. The
subsurface advantage is especially notable given that the forced challengers have
access to atmospheric information throughout the forecast that the emulator
does not. Two factors contribute. First, ongoing atmospheric forcing introduces
noise into the upper ocean that can amplify subsurface errors through mixed-layer
dynamics, whereas the emulator evolves a self-consistent water
column conditioned only on its initial state. Second, the architecture
operating on the native unstructured mesh preserves the topological relationships
between surface and subsurface nodes, which may help maintain vertical
consistency. Whether the remaining SST gap can be closed by re-introducing
atmospheric forcing as a conditioning input, without sacrificing the subsurface
and dynamical advantages, is an open question that motivates the forced extension
discussed in Section~\ref{sec:paths}.

\FloatBarrier
\section{Discussion}
\label{sec:discussion}

\subsection{The unstructured mesh}
\label{sec:mesh_discussion}
The results of Sections~\ref{sec:results} demonstrate
that running an ML emulator directly on the native unstructured FESOM2 mesh is
practical. Several specific findings support this claim.
 
First, the graph-based architecture treats the variable connectivity of the CORE2
mesh natively, and the forecast fields are spatially smooth across regions of very
different element size (Section~\ref{sec:skill}): neither the skill maps
(Fig.~\ref{fig:skill_maps}) nor the variance-ratio maps
(Fig.~\ref{fig:variability}) show discontinuities or degradation at
mesh-resolution transitions. Second, the spatial spectra computed directly on the
native triangular mesh (Fig.~\ref{fig:spectra}) exhibit the expected
scale-dependent behaviour without artifacts from regridding or interpolation.
Third, the OceanBench evaluation (Section~\ref{sec:oceanbench}) shows that the
same architecture, retrained on a regular-grid reanalysis product, produces
competitive or leading scores against systems that operate on regular grids,
confirming that the WeatherGenerator architecture does not sacrifice accuracy relative to
grid-based alternatives.
 
By itself, the CORE2 mesh is only mildly non-uniform (25--120\,km), and the
present results do not by themselves demonstrate an advantage over a regular grid of
comparable mean resolution. The value of the native-mesh approach becomes decisive
when the mesh is strongly refined. On the CORE2 mesh, the $\sim$127\,000 surface
vertices yield a graph roughly four times smaller than a uniform $1/4^\circ$ grid
($\sim$1\,million points), with proportional savings in memory and training cost.
The FESOM2 D3 configuration concentrates its $\sim$2\,million vertices in western
boundary currents and shelf seas, achieving $\sim$3\,km local resolution while
keeping the global vertex count an order of magnitude below that of a uniform
3\,km grid. The NG5 mesh reaches $\sim$5\,km globally. Both meshes represent the
state of the art in unstructured ocean modelling and are the natural next targets
for the emulator described here: the architecture requires no modification,
only retraining on the higher-resolution output.

\subsection{Comparison to other ocean ML systems}
\label{sec:comparison}
 
Table~\ref{tab:comparison} places the present model alongside the global ocean ML
systems reviewed in Section~\ref{sec:intro}. The comparison is necessarily
qualitative for three reasons: training data differ (coupled model output versus
reanalysis); verification targets differ (parent model versus reanalysis versus
observations); and evaluation protocols, lead times and variable sets are not
standardised across publications. The OceanBench intercomparison
(Section~\ref{sec:oceanbench}) provides one controlled comparison using a shared
protocol, but covers only the reanalysis-trained variant.
 
\begin{table}[htbp]
\centering
\caption{Qualitative comparison of global ocean ML systems. ``Forcing''
indicates whether the model receives atmospheric fields during the forecast rollout.
``Grid'' indicates the spatial discretisation. ``Depth'' indicates the number of
vertical levels. ``Training data'' indicates the
primary training target.}
\label{tab:comparison}
\small
\setlength{\tabcolsep}{4pt}
\begin{tabular}{l c c c c c}
\toprule
System & Grid & Forcing & Depth & Training data & Lead time \\
\midrule
HClimRep  & Unstructured & No & 20 & AWI-CM3 / GLORYS12 & 30\,d \\
XiHe \cite{xihe}      & Regular $1/12^\circ$ & Yes & 23 & GLORYS12 & 10--60\,d \\
WenHai \cite{wenhai}   & Regular $1/12^\circ$ & Yes & 23 & GLORYS12 & 10\,d \\
GLONET \cite{glonet}   & Regular $1/4^\circ$  & No  & 21 & GLORYS12 & 10\,d \\
LangYa \cite{langya}   & Regular $1/12^\circ$ & Yes & 32 & GLORYS12 & 7\,d \\
Njord \cite{njord}     & $1/4^\circ$ + $k$-means & Yes & 6 + sea ice & GLORYS12 (+GLO12) & 10\,d \\
FuXi-Ocean \cite{fuxiocean} & Regular $1/12^\circ$ & No & 20 & HYCOM & 10\,d \\
TianHai \cite{tianhai} & Regular $1/12^\circ$ & Yes (coupled) & 26 & HYCOM & 10\,d \\
Samudra \cite{samudra} & Regular $1^\circ$ & Yes & 19 (full) & OM4 output & Centuries \\
AI-GOMS \cite{aigoms}  & Regular $1/4^\circ$ & Yes & 15 & HYCOM & 30\,d \\
\bottomrule
\end{tabular}
\end{table}
 
Several observations emerge from the comparison. First, HClimRep is the only
system that operates on an unstructured mesh; all others either use a regular grid
directly or construct a graph by clustering a regular grid (Njord). Second, a
forcing-free rollout is not unique to HClimRep: GLONET and FuXi-Ocean also forecast
without explicit atmospheric inputs, relying on the recent ocean-state history to
carry the forcing signal implicitly. HClimRep differs in treating the forcing-free
protocol as a deliberate diagnostic of intrinsic ocean predictability rather than
an engineering simplification, and in extending the forecast horizon to 30 days.
The climate-emulation system Samudra, by contrast, is driven by surface fluxes and
targets centennial stability rather than short-range forecast skill. Third, the
systems that report the highest SST accuracy (WenHai, XiHe) both ingest atmospheric
forcing, which is consistent with the Hasselmann-framework interpretation: ongoing
atmospheric input is required to sustain the high-frequency surface variability
that determines pointwise SST skill. FuXi-Ocean is a partial exception, reporting
strong SST skill from ocean-only inputs, but it is trained and verified on six-hourly
HYCOM fields rather than GLORYS12 and is therefore not directly comparable. HClimRep's SST RMSD against the GLORYS
reanalysis (0.57--0.64\,\textdegree C over days~1--10;
Section~\ref{sec:oceanbench}) is competitive but not best-in-class, whereas its
subsurface temperature, salinity, current, SSH and Lagrangian scores are the lowest
reported in the intercomparison. This trade-off is a direct, physically motivated
consequence of the design rather than a deficiency.
 
A direct quantitative comparison of the FESOM2-trained emulator with the systems
in Table~\ref{tab:comparison} is not possible because they are verified against
different targets (the real ocean versus a coupled model's own trajectory).
The OceanBench variant bridges this gap by training on GLORYS12 and evaluating
under a shared protocol, confirming that the architecture and inference design
transfer to the observationally constrained setting.

\subsection{Limitations}
\label{sec:limitations}
The principal limitation is the forcing-free rollout. While it serves a
diagnostic purpose (Section~\ref{sec:horizon}), it prevents the emulator
from reproducing the oceanic response to atmospheric variability that occurs after
initialization. This accounts for the largest share of the spectral damping
documented in Section~\ref{sec:variability} and for the SST gap relative to
forced systems in the OceanBench evaluation.
 
The emulator is trained on a single realisation of the AWI-CM3 coupled model
under fixed greenhouse-gas forcing. It therefore inherits the biases of the
parent model, and verification against that same model
(Section~\ref{sec:results}) is not a test against the real ocean. The
OceanBench evaluation (Section~\ref{sec:oceanbench}) addresses this partly by
retraining on reanalysis, but the FESOM2-trained variant has not been evaluated
against observations.

The prognostic state is limited to 20 depth levels spanning the upper
$\sim$450\,m. The deep ocean is unconstrained, and processes that communicate
between the deep and upper ocean (e.g.\ deep-water formation, abyssal upwelling)
are not represented. Including deeper levels is straightforward in principle but
increases the graph size and training cost.

Finally, the model is a single deterministic forecast,
which by construction cannot represent forecast uncertainty or the full variance of
the flow.

\subsection{Paths to improvement}
\label{sec:paths}
The most direct remedy for the surface freezing is to re-introduce prescribed
atmospheric forcing during the rollout, restoring the high-frequency driver
(this requires extending the inference pipeline to accept step-wise atmospheric
fields, which it does not currently support). The
spectral bias of the deterministic, MSE-trained core can be attacked with
generative/probabilistic formulations (e.g.\ diffusion ensembles, in the spirit of
GenCast \cite{price2025gencast}), which sample rather than average and so preserve
variance, and with spectral or variance-aware loss terms that penalise the loss of
high-frequency power directly. Multi-step (rollout) training would expose the model
to its own compounding errors and curb the iterated-smoother effect, and training on
eddy-resolving rather than $\sim$1$^\circ$ data would supply the fine-scale
variability the current target lacks. By combining all those measures, one 
can expect model to perform skilled forecasts on multi-year scales. 

Training on eddy-resolving data is the route to representing the mesoscale. The
FESOM2 D3 mesh ($\sim$3\,km in western boundary currents) and NG5 mesh
($\sim$5\,km globally) are currently being integrated and will provide training
targets with an order of magnitude more spatial detail than the CORE2 output used
here.

\FloatBarrier
\section{Conclusions}
\label{sec:conclusions}
We have presented HClimRep-Ocean, to our knowledge the first ML ocean emulator
that runs directly on the native unstructured mesh of a global ocean model
(FESOM2). The architecture requires no regridding, shows no artifacts at
resolution transitions, and can be extended without modification to meshes with
$\sim$3--5\,km local resolution. A reanalysis-trained variant,
evaluated on OceanBench against five other forecasting systems, achieves the
lowest RMSD relative to GLORYS across the majority of variables, depths and lead
times, confirming that native-mesh emulation is competitive with regular-grid
approaches.
 
The forcing-free design reveals a clear, physically interpretable predictability
structure. Currents retain skill at all lead times and yield the smallest
Lagrangian trajectory deviations among all OceanBench challengers. SST and SSS
decay towards damped-anomaly persistence within roughly two weeks, as predicted
by the stochastic-climate framework of Hasselmann~(1976). Subsurface temperature
and salinity RMSD are lower than all other systems from 50 to 500\,m. The
pattern is not a deficiency but a diagnostic of where ocean predictability
resides when atmospheric forcing is removed.
 
Re-introducing atmospheric forcing is the most direct path to closing the SST
gap. Probabilistic formulations and eddy-resolving training data are
complementary next steps. The native-mesh foundation established here provides a
scalable basis for these extensions.

\FloatBarrier
\appendix

\section{Regional analysis}
\label{sec:regions}

The global scores of Sect.~\ref{sec:skill} average over regimes that behave very
differently. Here we repeat the verification within the six dynamically active regions
introduced in Sect.~\ref{sec:methods}, using the same 329 held-out initializations, the
same baselines and the same area weighting. Region boundaries are given in
Table~\ref{tab:regionbox} so that the numbers can be reproduced.

\begin{figure}[h]
  \centering
  \includegraphics[width=0.9\textwidth]{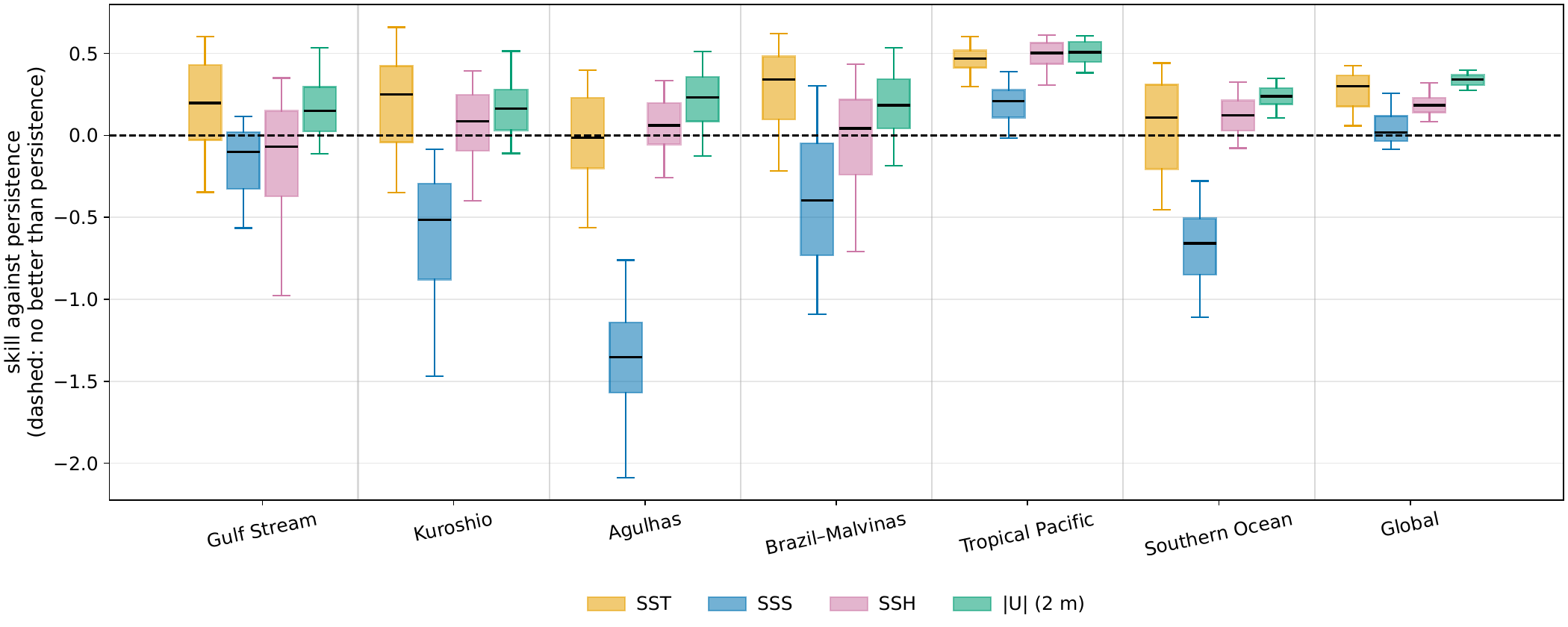}
  \caption{Distribution of the skill against persistence across 329 held-out
  initializations, by region and field, at 15-day horizon (box: interquartile range,
  whiskers: 5th to 95th percentile, line: median). Dashed lines mark parity with
  persistence. The distributions are strongly asymmetric --- skill cannot exceed one but is
  unbounded below --- and their width varies as much between regions as their mean does:
  in tropical Pacific every field is positive in $94$--$100\,\%$ of initializations,
  whereas in Gulf Stream surface temperature is positive in $71\,\%$ and sea-surface
  height in only $44\,\%$. Salinity in the Agulhas is negative in every single
  initialisation.}
  \label{fig:skillbox}
\end{figure}

Table~\ref{tab:regional} collects the results while in Figure~\ref{fig:skillbox} the corresponding distributions are depicted across initializations showing a left-skewed trend. Three patterns stand out. The tropical
Pacific is the best-forecast region by a wide margin and the only one in which every field,
salinity included, improves on persistence ($+0.20$ to $+0.51$ at 15 days); its variability
is dominated by comparatively deterministic equatorial and instability waves rather than by
a chaotic mesoscale. Surface currents, by contrast, improve on persistence everywhere
without exception ($+0.17$ to $+0.51$), making them the most uniformly skilful field
geographically as well as globally. Salinity is the mirror image: it is negative in every
region outside the tropical Pacific and worst in the Agulhas ($-1.36$) and the Southern
Ocean ($-0.67$), so the weakness identified globally in Sect.~\ref{sec:skill} is not spread
evenly but concentrated in the eddy-active regions of the southern hemisphere.

The two right-hand columns suggest why. The fraction of eddy kinetic energy the emulator
retains at 15 days tracks the regional skill closely: it is $0.64$ in the tropical Pacific,
where skill is highest, and only $0.17$--$0.25$ in the western boundary currents and the
Southern Ocean, where skill is lowest and salinity fails outright. At 107\,m the same
regions retain $0.62$--$0.85$, three to four times more than at the surface, consistent
with the depth dependence of the spectral damping documented in
Sect.~\ref{sec:variability}. Regional skill therefore appears to be set less by the
absolute variability of a region than by how much of its mesoscale the emulator is able to
carry. The global values in the last row illustrate the same point from the other side:
they are dominated by the quiescent open ocean and understate the deficit precisely where
the mesoscale matters most.

\begin{table}[b]
\centering
\caption{Region definitions. Longitudes are given in the range $-180^\circ$ to
$180^\circ$.}
\label{tab:regionbox}
\begin{tabular}{lcc}
\toprule
Region & Longitude & Latitude \\
\midrule
Gulf Stream       & $-80$ to $-40$   & $30$ to $50$   \\
Kuroshio          & $130$ to $170$   & $25$ to $45$   \\
Agulhas           & $10$ to $40$     & $-45$ to $-30$ \\
Brazil--Malvinas  & $-60$ to $-40$   & $-50$ to $-30$ \\
Tropical Pacific  & $-180$ to $-80$  & $-10$ to $10$  \\
Southern Ocean    & all              & $-65$ to $-45$ \\
\bottomrule
\end{tabular}
\end{table}

\begin{table}[t]
\centering
\caption{Regional skill at the 15-day horizon, measured against persistence as in
Table~\ref{tab:scores}. Skill is formed for each of the 329 held-out initialisations and
the table gives its mean $\pm$ the standard deviation across them; the standard error of
the mean is $0.02$ or smaller throughout, so the differences between regions are well
resolved. The last two columns give the median fraction of the reference eddy kinetic
energy retained at the same lead, at 2 and 107\,m, computed about the day-of-year
climatology over the same initialisations. Regions where the emulator retains more of the
mesoscale are also the regions in which it forecasts best.}
\label{tab:regional}
\begin{tabular}{lcccccc}
\toprule
 & \multicolumn{4}{c}{Skill vs persistence} & \multicolumn{2}{c}{Retained EKE} \\
\cmidrule(lr){2-5}\cmidrule(lr){6-7}
Region & SST & SSS & SSH & $|U|$ & 2\,m & 107\,m \\
\midrule
Gulf Stream       & $+0.17{\pm}0.32$ & $-0.15{\pm}0.22$ & $-0.15{\pm}0.43$ & $+0.17{\pm}0.20$ & $0.17$ & $0.62$ \\
Kuroshio          & $+0.19{\pm}0.31$ & $-0.62{\pm}0.44$ & $+0.06{\pm}0.24$ & $+0.17{\pm}0.19$ & $0.23$ & $0.79$ \\
Agulhas           & $-0.02{\pm}0.30$ & $-1.36{\pm}0.40$ & $+0.06{\pm}0.19$ & $+0.21{\pm}0.21$ & $0.25$ & $0.68$ \\
Brazil--Malvinas  & $+0.28{\pm}0.27$ & $-0.40{\pm}0.44$ & $-0.03{\pm}0.37$ & $+0.19{\pm}0.23$ & $0.22$ & $0.65$ \\
Tropical Pacific  & $+0.46{\pm}0.09$ & $+0.20{\pm}0.13$ & $+0.49{\pm}0.10$ & $+0.51{\pm}0.08$ & $0.64$ & $0.85$ \\
Southern Ocean    & $+0.05{\pm}0.30$ & $-0.67{\pm}0.26$ & $+0.12{\pm}0.13$ & $+0.24{\pm}0.07$ & $0.24$ & $0.67$ \\
\addlinespace
\textit{Global}   & $+0.27{\pm}0.12$ & $+0.05{\pm}0.11$ & $+0.19{\pm}0.07$ & $+0.34{\pm}0.04$ & $0.50$ & $0.81$ \\
\bottomrule
\end{tabular}
\end{table}

The standard deviations are large in the eddy-active regions: surface-temperature skill in
the Gulf Stream is $+0.17\pm0.32$, so it is negative for roughly a third of the
initialisations, whereas in the tropical Pacific it is $+0.46\pm0.09$ and positive for
essentially all of them (Fig.~\ref{fig:skillbox}). The regions in which the emulator
performs best are therefore also the regions in which it performs most consistently.

\begin{figure}[h]
  \centering
  \includegraphics[width=\textwidth]{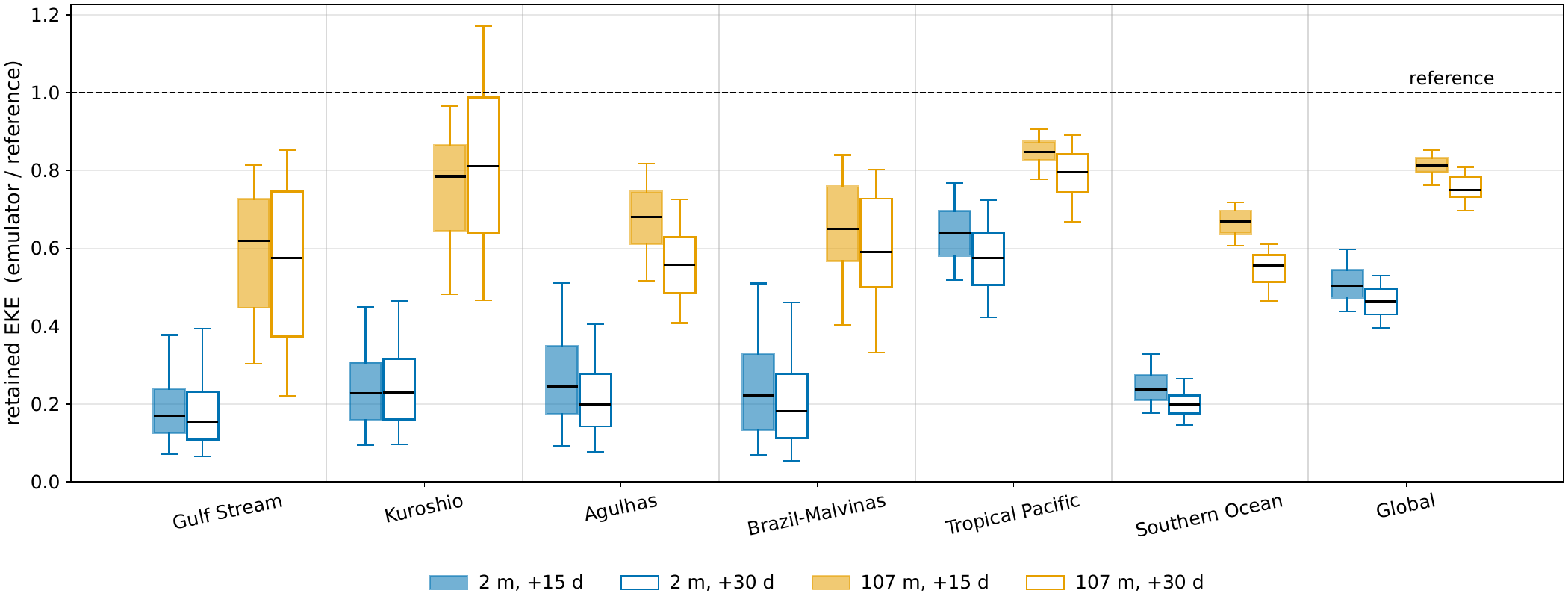}
  \caption{Fraction of the reference eddy kinetic energy retained by the
  emulator, by region. Eddy kinetic energy is formed about the day-of-year climatology and
  averaged within each region with the FESOM2 element areas; each box shows the
  distribution over the 329 held-out initialisations (box: interquartile range, whiskers:
  5th to 95th percentile, line: median). Colour denotes depth and fill denotes lead time.
  The deficit is severe at the surface in the eddy-active regions --- a median of $0.17$ in
  the Gulf Stream against $0.64$ in the tropical Pacific --- and three to four times
  smaller at 107\,m. It is also established early: the medians change little between +15
  and +30 days, except in the Agulhas and the Southern Ocean, where the deficit continues
  to deepen. The spread between initialisations is itself informative: in the western
  boundary currents the retained fraction ranges from below $0.1$ to above $0.4$ depending
  on the start date, whereas the tropical Pacific and the Southern Ocean behave
  consistently. The Kuroshio at 107\,m is the only case in which the emulator exceeds the
  reference, carrying more deep eddy energy than it should in $24\,\%$ of initialisations
  at +30 days.}
  \label{fig:ekebox}
\end{figure}

\newpage

\begin{table}[h]
\centering
\caption{Training stages. Each stage is initialised from the preceding checkpoint;
``steps'' are optimiser steps at a global batch of eight rollouts, the horizon is given
in daily forecast steps and $\eta_{\max}$ is the peak learning rate.}
\label{tab:app_stages}
\begin{tabular}{lccll}
\toprule
Stage & Horizon & $\eta_{\max}$ & Trainable weights & Steps \\
\midrule
Pre-training & 4 & $5\times10^{-5}$ & all ($1.138\times10^{9}$) & $\sim$32\,900 \\
Forecast fine-tuning & 15 & $10^{-5}$ & forecasting engine & $\sim$6\,700 \\
Final refinement & 15 & $10^{-5}$ & global aggregation engine & 1\,408 \\
\bottomrule
\end{tabular}
\end{table}

\begin{table}[t]
\caption{Model configuration of HClimRep-Ocean prototype model.}
\label{tab:config}
\centering
\begin{tabular}{p{5cm} p{7cm} c}
\hline
\textbf{Model Component} & \textbf{Configuration} & \textbf{Value} \\
\hline
\hline
Embedding & embed\_unembed\_mode & block \\
 & HEALPix level & 5 \\
 & Dropout rate & 0.1 \\
\hline
Local Assimilation Engine & Attention blocks & 4 \\
 & Attention heads & 16 \\
 & Embedding dimension & 2048 \\
 & Dropout rate & 0.1 \\
\hline
Local Adapter & Attention blocks & 2 \\
 & Attention heads & 16 \\
 & Embedding dimension & 2048 \\
 & Dropout rate & 0.1 \\
\hline
Global Assimilation Engine & Attention blocks & 12 \\
 & Attention heads & 32 \\
 & Embedding dimension & 2048 \\
 & Dropout rate & 0.1 \\
\hline
Decoder & Attention blocks & 2 \\
 & Attention heads & 4 \\
 & Embedding dimension & 256 \\
 & Dropout rate & 0.1 \\
\hline
Forecasting Engine & Attention blocks & 16 \\
 & Attention heads & 16 \\
 & Dropout rate & 0.1 \\
\hline
\end{tabular}
\end{table}

\section{Training setup}
\label{app:training}

\subsection{Model Architecture}
\label{app:model_architecture}

\subsection{Optimization}
\label{app:optim}

The model has $1.138\times10^{9}$ parameters. All stages are trained with AdamW \cite{adamw} using
$\beta_1=0.98125$, $\beta_2=0.9875$, $\varepsilon=2\times10^{-8}$ and a weight decay of
$0.1$, with gradients clipped to a global norm of $1.0$. The learning rate follows a
three-phase schedule: a cosine warm-up over 256 steps from $10^{-6}$ to the stage
maximum, a cosine decay towards $2\times10^{-6}$, and a linear cool-down to zero over the
final 512 steps. The maximum learning rate is $5\times10^{-5}$ for pre-training and
$10^{-5}$ for all fine-tuning stages, and is rescaled with the square root of the number
of ranks. Training uses bfloat16 mixed precision with bfloat16 attention and
FlashAttention, distributed over eight NVIDIA A100 GPUs (two nodes) with combined DDP and
FSDP; \texttt{torch.compile} is disabled. One sample is processed per GPU, giving a global
batch of eight complete forecast rollouts per optimizer step, with eight data-loading
workers per rank and a fixed random seed.

The objective is a physical-space loss combining a mean-squared-error term with a dynamic
loss that renormalises each channel by a running estimate of its error scale (window 128,
$L=20$), which prevents the numerically large temperature and salinity channels from
dominating the small velocity and sea-level channels. Both prognostic streams enter the
loss with unit weight; the atmospheric stream is diagnostic and contributes no loss term.

\subsection{Training protocol}
\label{app:protocol}

The model was trained in stages rather than in a single run, each stage initialized from
the preceding checkpoint (Table~\ref{tab:app_stages}). Pre-training used a masked-token
objective with a four-step forecast horizon and no frozen weights. The forecast
fine-tuning stage extended the horizon to the 15 steps used at inference while freezing
the encoder's local, global and adapter blocks together with the stream embeddings, so
that only the forecasting engine was updated. 
The final stage refined the model global aggregation engine with $268.5$\,M parameters 
($23.6$\,\% of the total) left trainable and the
remainder frozen, for 1408 optimiser steps in approximately twelve hours of wall-clock
time.

Training data are daily fields from the AWI-CM3 FESOM2 simulation covering the years
2000--2208; the following year is held out entirely and provides the initialisations used
in Sect.~\ref{sec:results}. Each mini-epoch draws 2048 training samples (4096 during
pre-training) and validation during training used 64 samples per mini-epoch from the
held-out year, without exponential moving averaging of the weights.

\section*{Author contributions}
KN designed the study, trained the emulator, carried out the forecast experiments, prepared evaluation, produced the figures and wrote the manuscript. AK contributed to data preparation, carried out data analysis and contributed to writing the manuscript. NK prepared the simulation output used for training and verification. SM, AP, SG and JP contributed to software development. SM contributed on writing the manuscript. MS, CL, SM and JP reviewed the manuscript. CL, MS and TJ supervised the project and acquired funding. All authors read and approved the final manuscript.

\section*{Acknowledgements}

The authors gratefully acknowledge the computing resources provided on the high
performance computer Levante at the German Climate Computing Center (DKRZ).

The authors thank the Gauss Centre for Supercomputing e.V. (GCS) for providing computing time on the Supercomputer JUPITER at the Jülich Supercomputing Centre. JUPITER is supported by the EuroHPC JU and GCS through funding by the European Commission, the German Federal Ministry of Research, Technology and Space, and the Ministry of Culture and Science of the State of North Rhine-Westphalia.

This work used data produced by the EERIE project, funded by the European Union's Horizon Europe programme under grant agreement No 101081383.

NK was also supported by the CLAIMA project (grant agreement No 101290971), funded by the European Union.

This work was supported by the Helmholtz Association within the framework of the Helmholtz Foundation Model Initiative (HClimRep).

Finally, the authors would like to thank the WeatherGenerator and HClimRep consortia and their developers for providing the open-source code used in this work. We are also grateful for their collaborative support and technical and scientific assistance. 

The WeatherGenerator project (grant agreement No. 101187947) is funded by the European Union. Views and opinions expressed are however those of the author(s) only and do not necessarily reflect those of the European Union or the European Research Executive Agency (REA). Neither the European Union nor the granting authority can be held responsible for them.

\bibliographystyle{unsrt}
\bibliography{refs}

\end{document}